\documentclass[11pt]{article}
\usepackage{amsmath}
\usepackage{latexsym}
\usepackage{graphicx}
\usepackage{bbm}
\usepackage{color}
\usepackage{euscript}
\usepackage{amsfonts}
\usepackage{cite}

\makeatletter
\@addtoreset{equation}{section}
\makeatother

\def\beq{\begin{equation}}
\def\eeq{\end{equation}}
\def\be{\begin{equation}}
\def\ee{\end{equation}}

\def\bea{\begin{eqnarray}}
\def\eea{\end{eqnarray}}
\def\ft#1#2{{\textstyle{\frac{\scriptstyle #1}{\scriptstyle #2}}}}
\def\fft#1#2{\frac{#1}{#2}}

\let\ep=\epsilon

\let\m=\mu 
\let\n=\nu

\let\bm=\bibitem

\newcommand{\tr}{{\rm tr} }

\def\R{{\mathbbm R}}

\def\Z{{\mathbbm Z}} 

\def\cK{{\cal K}}
\def\cG{{\cal G}}
\def\cH{{\cal H}}
\def\cV{{\cal V}}

\def\cJ{{\cal J}}
\def\bG{{\bf G}}
\def\bH{{\bf H}}
\def\bK{{\bf K}}
\def\del{{\partial}}
\def\bm{\bibitem}
\def\sst#1{{\scriptscriptstyle #1}}
\def\n{{\sst (n)}}
\def\m{{\sst (m)}}
\def\p{{\sst (p)}}
\def\nn{\nonumber}
\def\td{\tilde}
\def\wtd{\widetilde}
\def\bep{{\bar\ep{\,}}}
\def\ie{{\it i.e.\ }} 

\newcount\hour \newcount\minute
\hour=\time  \divide \hour by 60
\minute=\time
\loop \ifnum \minute > 59 \advance \minute by -60 \repeat
\def\nowtwelve{\ifnum \hour<13 \number\hour:
                      \ifnum \minute<10 0\fi
                      \number\minute
                      \ifnum \hour<12 \ A.M.\else \ P.M.\fi
	 \else \advance \hour by -12 \number\hour:
                      \ifnum \minute<10 0\fi
                      \number\minute \ P.M.\fi}
\def\nowtwentyfour{\ifnum \hour<10 0\fi
		\number\hour:
         	\ifnum \minute<10 0\fi
         	\number\minute}

\begin{document}

\voffset=0.3truein
\hfuzz=100pt

\title{Infinite-Dimensional Symmetries of Two-Dimensional Coset Models}
\author{\Large H. L\"u,$^1$ Malcolm J. Perry$^2$ and
C.N. Pope$^{1,2}$
\\ \\ \\
${}^1$ George P. \& Cynthia W. Mitchell Institute for Fundamental Physics,\\
         Texas A\&M University,
         College Station,
         TX 77843-4242,
         USA.\\ \\
${}^2$ DAMTP, Centre for Mathematical Sciences,\\
         University of Cambridge,
         Wilberforce Road,
         Cambridge CB3 0WA,
         England.\\ \\ \\  \\} 

\date{\empty}
\maketitle

\includegraphics[scale = 0.05, bb= 0 -6800 100 -8600 ]{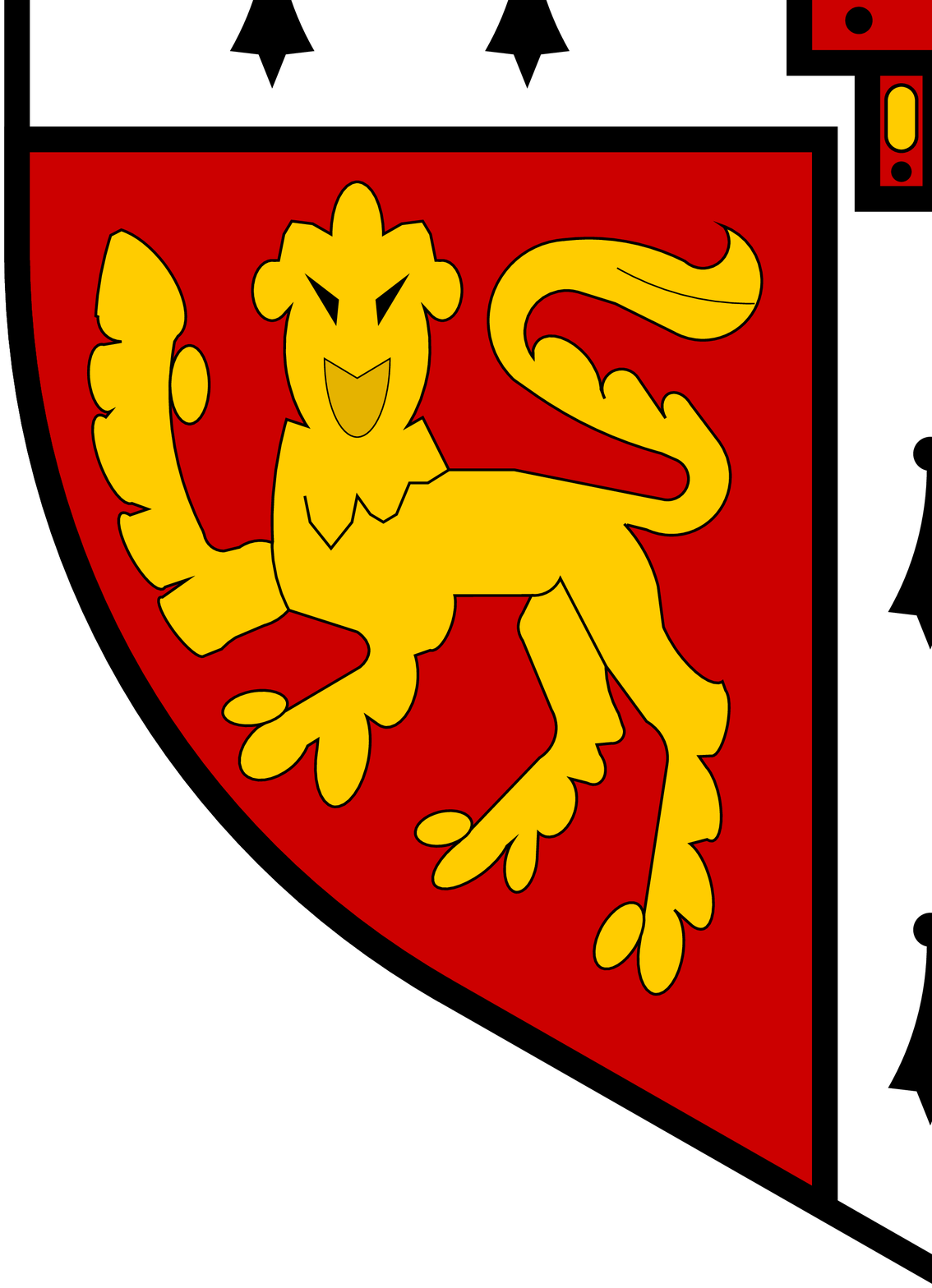}
\includegraphics[scale = 0.17, bb= -1900 -2000 0 -1800 ]{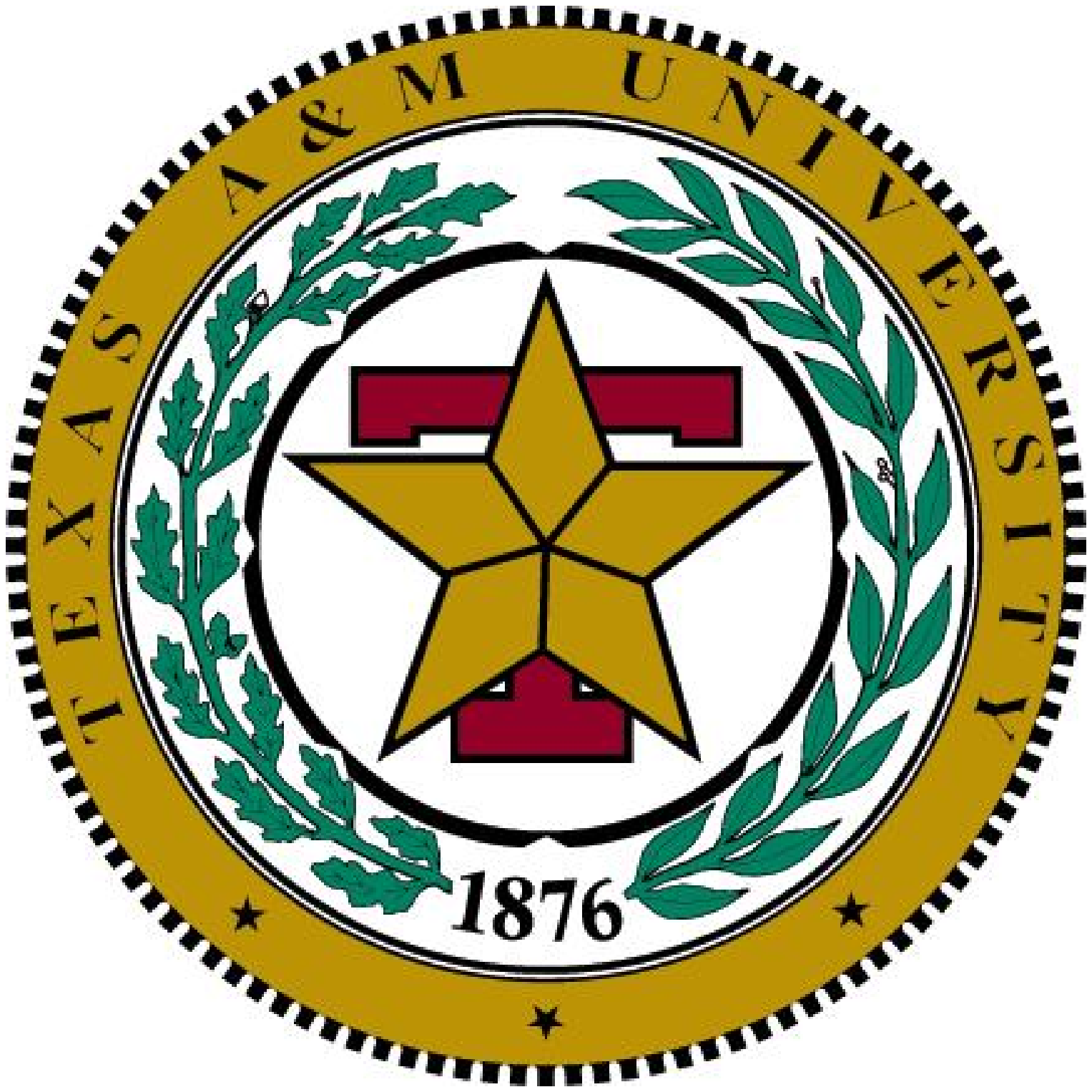}

\vskip -5.5truein

\centerline{DAMTP-2007-107\hskip 1.5truein MIFP-07-28}
\bigskip\bigskip
\centerline{\bf{ArXiv:0711.0400}}

\vskip 3.5truein


\begin{abstract} 

  It has long been appreciated that the toroidal reduction of any gravity or 
supergravity to two dimensions gives rise to a scalar coset theory exhibiting
an infinite-dimensional global symmetry.  This symmetry is an 
extension of the finite-dimensional symmetry $\cG$ in three dimensions, after
performing a further circle reduction.  There has not been 
universal agreement as to exactly what the extended symmetry algebra is,
with different arguments seemingly concluding either that it is $\hat \cG$,
the affine Kac-Moody extension of $\cG$, or else a subalgebra thereof. 
We take the very explicit approach of Schwarz as our 
starting point for studying the
simpler situation of two-dimensional flat-space sigma models, which nonetheless
capture all the essential details.  We arrive at the conclusion that 
the full symmetry is described by the Kac-Moody algebra $\hat\cG$, whilst
the subalgebra obtained by Schwarz arises as a gauge-fixed truncation.
We then consider the explicit example of the $SL(2,\R)/O(2)$ 
coset, and relate Schwarz's approach to an earlier discussion that goes
back to the work of Geroch.

\end{abstract}

\thispagestyle{empty}

\pagebreak
\voffset=0pt
\setcounter{page}{1}

\tableofcontents

\addtocontents{toc}{\protect\setcounter{tocdepth}{2}}

\def\V{\mathcal{V}}
\def\hV{\hat{\mathcal{V}}}


\section{Introduction}

   The study of supergravity theories, and their symmetries, have played
a very important r\^ole in uncovering the underlying structures of
string theory.  Especially significant are the U-duality symmetries of
the string, which have their origin in the classical global symmetries
exhibited by eleven-dimensional supergravity and type IIA and IIB 
supergravities
after toroidal dimensional reduction.  For example, if one reduces 
eleven-dimensional supergravity on an $n$-torus, for $n\le8$, the
resulting $D=(11-n)$-dimensional theory exhibits a global $E_{n}$ symmetry
\cite{crejul,cj2,cjlp}.  In the cases $n\ge3$ this symmetry arises in quite a
subtle way, involving an interplay between the original eleven-dimensional
metric and the 3-form potential.

   In view of the large $E_8$ symmetry that one finds after reduction to
three dimensions, it is natural to push further and investigate the symmetries
after further reduction to two dimensions, and even beyond.  It turns out that
the analysis of the global symmetry 
for a reduction to two dimensions is considerably more 
complicated than the higher-dimensional ones.  There are two striking new 
features that lead to this complexity.  The first is that, unlike a 
reduction to $D\ge3$ dimensions, one can no longer use a reduction 
scheme in which the metric is reduced from an Einstein-frame metric in the
higher dimension to an Einstein-frame metric in the lower dimension.  (In
the Einstein conformal frame, the Lagrangian for gravity itself takes the 
form ${\cal L} \sim \sqrt{-g} R$, with no scalar conformal factor.)  The
inability to reach the Einstein conformal frame in two dimensions is 
intimately connected to the fact that $\sqrt{-g} R$ is a conformal
invariant in two dimensions.  It has the consequence that the metric in
two dimensions is not invariant under the global symmetries.

   The second striking new feature is that an axionic
scalar field (\ie a scalar appearing everywhere covered by a derivative)
can be dualised to give another axionic scalar field in the
special case of two dimensions.  This has the remarkable consequence that 
the global symmetry group actually becomes infinite in dimension.  This
was seen long ago by Geroch, in the context of four-dimensional gravity
reduced to two.   There are degrees of freedom
in two dimensions that are described by the sigma model $SL(2,\R)/O(2)$,
and under dualisation this yields another $SL(2,\R)/O(2)$ sigma model.
Geroch showed that the two associated global $SL(2,\R)$ symmetries do
not commute, and that if one takes repeated commutators of the two
sets of transformations, an infinite-dimensional algebra results \cite{geroch}.
The precise nature of this symmetry, now known as the {\it Geroch Group},
was not uncovered in \cite{geroch}. 

   The feature of having an infinite-dimensional symmetry in two 
dimensions is not restricted to situations where gravity is involved,
and in fact the same essential mechanism operates in a similar fashion
if one considers a sigma model in a flat two-dimensional spacetime.  Thus,
a natural preliminary to investigating the symmetries of two-dimensional
reductions in supergravity is to study the symmetry of a flat two-dimensional
sigma model $\bG/\bH$.  Considerable simplifications arise if one restricts
attention to symmetric-space sigma models, and since these in any 
case always
arise in supergravity dimensional reductions, the specialisation to this
class of models is a very natural one.  We shall use the acronym SSM to 
denote a symmetric-space sigma model.

   There is quite a considerable literature on the subject of the infinite
dimensional symmetries of two-dimensional symmetric-space sigma models,
both in the flat and the curved spacetime cases (see, for example, 
\cite{pol,lus,polus,kinchi,belzak,mais1,breitmais,dolan,devfair,yswu,jul},
some of which considers also principal chiral models).  
A very clear and explicit presentation of the global symmetry algebras of 
two-dimensional SSMs has been provided by
Schwarz, whose papers formulate the problem in
a very transparent way.  He first considers the problem of two-dimensional
theories in flat spacetime in \cite{schwarz1}, and then generalises to the
case of a curved two-dimensional spacetime in \cite{schwarz2}.  
He also gives an extended history of the earlier
literature, and rather than attempting to repeat that here, we refer the
reader to his papers for further details.

   Our work in the present paper is concerned entirely with the case
of symmetric-space sigma models in flat two-dimensional spacetime, and
we follow very closely the approach taken by Schwarz in \cite{schwarz1}.  
The results in \cite{schwarz1} differ somewhat from those in much of the
literature, where the infinite-dimensional
global symmetry algebra of the SSM $\bG/\bH$ is found to be $\hat\cG$, 
the affine Kac-Moody
extension of the underlying algebra $\cG$ of the ``manifest'' global symmetry
group $\bG$.  The generators of $\hat\cG$ may be represented by $J^i_n$,
satisfying
\be
[J^i_m,J^j_n]= f^{ij}{}_k\, J^k_{m+n}\,,
\ee
where $f^{ij}{}_k$ are the structure constants for the Lie algebra
$\cG$, whose generators $T^i$ satisfy $[T^i,T^j]=f^{ij}{}_k\, T^k$.

   By contrast, Schwarz obtained a certain subalgebra $\hat\cG_H$ of 
$\hat\cG$ as the global symmetry algebra,
essentially generated by ${J'}^i_n= J^i_n\pm J^i_{-n}$, where the $+$
sign is chosen if $i$ lies in the denominator algebra $\cH$, and the $-$
sign if $i$ lies in the coset $\cK=\cG/\cH$.

   We find that by extending the techniques developed by Schwarz, we can
construct explicit global symmetries for the entire
$\hat \cG$ Kac-Moody algebra, expressed purely in terms of local field
transformations.  As far as we are aware, it is only through
the use of the construction that Schwarz developed that it has become 
possible to obtain explicit local transformations for the entire Kac-Moody
algebra.  We find also that the subalgebra obtained by Schwarz can 
be viewed as a gauge-fixed version of this full Kac-Moody symmetry. 

     In order to understand this, we recall that the possibility of 
dualising axions to give new axions in two dimensions means that the 
original theory can be reformulated in terms of new fields that are
non-locally related to the original ones (since the process of dualisation
requires differentiation and Hodge dualisation, followed by integration,
to obtain the new variables). A convenient
way to handle this is to enlarge the system by introducing auxiliary
fields, so that the manifest global symmetries of the original and the
dualised sigma models can be exhibited simultaneously, in purely local terms. 
In fact to do this, one has to introduce an infinite number of auxiliary 
fields.  
The full set of Kac-Moody symmetries, generated by
$J^i_n$ with $-\infty\le n\le \infty$, acts on the complete set of original
plus auxiliary fields. However, the ``negative
half'' of the Kac-Moody algebra $\hat \cG$, generated by $J^i_n$ with $n<0$, 
acts exclusively on the auxiliary fields, whilst leaving the original
sigma-model fields inert.  In fact these symmetries are essentially
constant shift transformations of the auxiliary fields, reflecting the
arbitrariness of the choice of constants of integration that arose when
the non-local dualisation was recast into a local form in terms of the
auxiliary fields. 
   
   The subalgebra $\hat\cG_H$ of symmetries found by Schwarz can be
viewed as a gauge fixing in which the values of the original and the
auxiliary fields are all set to prescribed values at some chosen point
in the two-dimensional spacetime.  Effectively, the level-0 transformations
$J^i_0$ that lie in $\cK$ are used up in gauge fixing the original fields to
their prescribed values, and the entirety of the $J^i_n$ transformations
with $n<0$ are used up in doing the same for the auxiliary fields.

   The formulation in which the auxiliary fields
are added has been developed considerably by Nicolai, and Julia 
\cite{nicolai,nicjul}.  However, the work of
Schwarz provided a procedure for obtaining explicit expressions 
for the transformations
associated with the ``upper half'' of the Kac-Moody algebra. We are 
able to draw the two approaches together and provide a fully explicit
and local description of the entire Kac-Moody algebra of symmetries.

   In order to illustrate these ideas in detail, it is useful to
examine an example.  For this purpose we choose the simplest non-trivial
symmetric-space sigma model, $SL(2,\R)/O(2)$.  We show how one needs
to introduce an infinity of auxiliary fields in order to describe 
simultaneously the original $SL(2,\R)$ symmetry and the $SL(2,\R)$
symmetry of the dualised version (we denote this by $\overline{SL(2,\R)}$).
We also show how each generator of each copy of $SL(2,\R)$ can be precisely
matched with a corresponding generator in the Kac-Moody algebra, and this
allows us to show explicitly that the Geroch algebra generated by taking
multiple commutators of $SL(2,\R)$ and $\overline{SL(2,\R)}$ transformations
is exactly the same as the full Kac-Moody algebra $\widehat{SL(2,\R)}$.

  We also examine a further symmetry of two-dimensional symmetric-space
sigma models $\bG/\bH$, again
basing our analysis on the work of Schwarz \cite{schwarz1}.  This is again
an infinite-dimensional symmetry, but this time a singlet under the
original $\cG$ symmetry.  It turns out to be related to the centreless
Virasoro algebra.

\section{Lax Equation and Infinite-Dimensional Symmetries}

\subsection{Basic formalism}\label{formsec}

   We shall begin by considering an arbitrary symmetric-space sigma model 
(SSM) in a flat two-dimensional spacetime background, with coset 
given by $\bK=\bG/\bH$,
where $\bG$ is a Lie group with subgroup $\bH$.  The commutation relations
for the corresponding generators of the algebra take the form
\be
[\cH,\cH]=\cH\,,\qquad [\cH,\cK]=\cK\,,\qquad
   [\cK,\cK]=\cH\,.\label{coset}
\ee
The condition that $\bK$ is a symmetric space is reflected in the absence of
$\cK$ generators on the right-hand side of the last commutation relation.
The symmetric-space algebra implies that there is an involution $\sharp$ 
under which
\be
\cK^\sharp = \cK\,,\qquad \cH^\sharp=-\cH\,.
\label{invol1}
\ee
In many cases, such as when $\bG=SL(n,\R)/O(n)$, the involution map is given
by Hermitean conjugation,
\be
\cK^\dagger=\cK\,,\qquad \cH^\dagger
 = -\cH\,,\label{invol2}
\ee
and later, we shall typically write formulae under this assumption.  
In some cases,
such as $\bG=E_{(8,8)}$, $\cH=O(16)$, the involution $\sharp$ is more involved. 

  Let $\cV$ be a coset representative in $\bK$.  We may then define
\be
M=\cV^\sharp\, \cV\,,\qquad A= M^{-1} dM\,.\label{MAdef}
\ee
Under transformations 
\be
\cV\longrightarrow h\cV g\,,\label{gtrans}
\ee
where $g$ is a global element
in the group $\bG$ and $h$ is a 
local element in the denominator subgroup $\bH$, we have shall have
\be
M\longrightarrow g^\sharp M g\,,\qquad 
                   A\longrightarrow g^{-1} A g\,,\label{Mtrans}
\ee
since it follows from $\cH^\sharp =-\cH$ that $h^\sharp= h^{-1}$.

   The Cartan-Maurer equation $d(M^{-1} dM)= -(M^{-1} dM)\wedge (M^{-1} dM)$
implies that the field strength for $A$ vanishes:
\be
F\equiv dA + A\wedge A=0\,.\label{cartmau}
\ee
The Lagrangian for the coset model may be written as $L= -\ft14 
\tr({*A}\wedge A)$ (or,
using indices, $L=- \ft14 \eta^{\mu\nu} \tr(A_\mu A_\nu)$) and 
hence the equation of motion is
\be
 d{*A}=0\,.\label{eom}
\ee
The Lagrangian is clearly invariant under the global $\bG$ transformations,
and the equations (\ref{cartmau}) and (\ref{eom}) transform covariantly under
$\bG$.

   As discussed in \cite{schwarz1}, the equations (\ref{cartmau}) 
and (\ref{eom}) 
can both be derived from the integrability condition for the 
{\it Lax Pair} of linear equations
\be
 \Big(\del_+ + \fft{t}{t-1} A_+\Big) X=0\,,\qquad
 \Big(\del_- + \fft{t}{t+1} A_-\Big) X=0\,,\label{Lax1}
\ee
to admit a solution $X(x;t)$,
where $t$ is an arbitrary constant {\it spectral parameter}.  
These equations are written in
light-cone coordinates on the two-dimensional flat spacetime, in which the 
metric is $ds^2=2 dx^+ \, dx^-$.  We prefer to use the language of differential
forms, for which $A= A_+ dx^+ + A_- dx^-$.  On 1-forms we have $*^2=+1$, where
$*$ is the Hodge dual operator, and 
\be
{*dx^\pm} = \pm dx^\pm\,,\qquad \hbox{and so}\qquad {*A}=A_+ dx^+ - A_- dx^-\,.
\ee
It is useful also to record the following properties for 1-forms $u$ and $v$:
\be
{*u}\wedge v= {*v}\wedge u\,,\qquad 
             {*u}\wedge {*v}= - u\wedge v\,,\label{uvrels}
\ee
and for Lie-algebra valued 1-forms $A$ and $B$:
\be
{*A}\wedge B =- A\wedge {*B}\,,\qquad 
               {*A}\wedge {*A} =- A\wedge A\,.\label{ABrels}
\ee

   In terms of differential forms, the Lax pair (\ref{Lax1}) becomes simply the
single equation
\be
t (d+A) X= {*dX}\,.\label{Lax2}
\ee
We shall call this the {\it Lax Equation}.  By taking the appropriate linear
combination of this and its dual, we obtain
\be
dX X^{-1} = \fft{t}{1-t^2} \, {*A} + \fft{t^2}{1-t^2} \, A\,.\label{Lax3}
\ee
Thus the integrability condition for the existence of a solution 
$X(x;t)$ to the
Lax equation, which follows from the Cartan-Maurer equation $d(dX X^{-1})=
(dX X^{-1})\wedge (dX X^{-1})$, gives
\be
d{*A} + t (dA + A\wedge A)=0\,.
\ee
Since this must hold for all $t$ we indeed derive (\ref{cartmau}) and
(\ref{eom}).  Note that (\ref{Lax3}) is an equivalent formulation of the 
Lax equation; an appropriate linear combination of (\ref{Lax3}) and
its dual gives back (\ref{Lax2}).  Thus we may use the term ``Lax equation''
interchangeably for (\ref{Lax2}) and (\ref{Lax3}).

\subsection{Infinite-dimensional extension of the global $\bG$ symmetry}
\label{kacmoodysec}

   We have already noted that the global $\bG$ transformations (\ref{Mtrans}) 
are a symmetry of the zero-curvature condition (\ref{cartmau}) and the
equations of motion (\ref{eom}) of the two-dimensional coset model.  In fact,
these symmetries are merely the tip of an infinite-dimensional ``iceberg'' of
global symmetries.  These extended symmetries are a special feature that arises
because the coset model lives in a two-dimensional world volume, and they may
be understood in a variety of ways.  An intuitive understanding, which we
shall turn into a concrete discussion in section \ref{sl2rsec} for the 
example of
the coset $SL(2,\R)/O(2)$, is that the axionic scalars can be dualised into
new, non-locally related sets of axions in two dimensions, and that the
manifest global symmetries in the different duality pictures do not commute,
but instead their commutators close only on an infinite-dimensional 
extension of the finite-dimensional
symmetries that are manifest in each individual duality choice.\footnote{This
idea dates back to a paper on four-dimensional gravity reduced
to two dimensions, by Geroch \cite{geroch}, although at that time the
precise nature of the infinite-dimensional algebra was not addressed.}

   In the 
present section, we shall begin by following a construction given in 
\cite{schwarz1}, which shows how the formalism of the Lax equation may be
used to derive the infinite-dimensional algebra.  Our description will be
formulated in the language of differential forms rather than light-cone
coordinates.  The details of our calculation differ somewhat from those
in \cite{schwarz1}, and our conclusions differ also.  Specifically, we
find that the full symmetry of the symmetric-space sigma model is
precisely the affine Kac-Moody extension $\hat\cG$ of the manifest
$\cG$ global symmetry, and not merely the subalgebra of $\hat\cG$ that
was found in \cite{schwarz1}.  (We shall comment further about this 
later in this subsection, and in appendix \ref{appA}.)

   At the infinitesimal level, the transformation (\ref{gtrans}) becomes
\be
\delta \cV = \cV \ep + \delta h\, \cV\,,
\ee
where $\ep$ is an infinitesimal 
global element of the Lie algebra $\cG$ and $\delta h$ is
a local element of $\cH$.  In order to exhibit the infinite-dimensional
extension of this symmetry algebra, one may consider more general 
transformations of the form \cite{schwarz1}
\be
\delta \cV = \cV \eta + \delta h \, \cV\,,\qquad \hbox{where}\qquad \eta=
        X(t) \ep X(t)^{-1}\,.\label{deltag}
\ee
The meaning of this equation is as follows.  As before, $\cV$ is a coset
representative for $\bG/\bH$, and thus it depends on the scalar fields 
parameterising the coset, which themselves depend on the two spacetime
coordinates $x$, but it does not depend on the spectral parameter $t$.  
The function $X(t)$ is the solution of the Lax equation
(\ref{Lax2}) and thus it depends on the spacetime coordinates $x$ (we are
now suppressing the explicit indication of this dependence) and on the spectral
parameter $t$.  The quantity $\delta h$, in the denominator algebra $\cH$, is
a function of the spacetime fields and it may now depend upon $t$.  On
the left-hand side of (\ref{deltag}) there is $t$-dependence only in the
variational symbol $\delta$ itself, and it is to be interpreted as
\be
\delta =\delta(\ep,t) = \sum_{n\ge 0} t^n\, \delta_\n(\ep)\,.\label{deltan}
\ee
Thus by equating powers of $t$ on the two sides of (\ref{deltag}) we obtain a
hierarchy of transformations $\delta_\n$ that act upon the scalar fields in
the coset representative $\cV$.  The lowest set of transformations, \ie for
$n=0$, just correspond to the original infinitesimal $\bG$ transformations
that were manifest in the coset model from the outset.  By contrast
the transformations $\delta_\n$ with $n>0$, which all involve $t$-dependent
terms in $X(t)$, are non-local expressions in terms of the original fields of
the scalar coset.\footnote{Note, however, that all the transformations 
become local if one introduces an infinite set of auxiliary fields, as
we shall do later.} 

   To check that (\ref{deltag}) does indeed give symmetries of the theory,
one must check that the corresponding variation of the equation of motion 
(\ref{eom}) vanishes.  First, one sees from\footnote{From this point
onwards, we shall assume for simplicity, and to make the expressions look 
more palatable, that the involution of the symmetric space algebra is 
implemented by Hermitean conjugation, as in (\ref{invol2}).  In a case 
where the more general
$\sharp$ involution operator is required, all $\dagger$ symbols in what follows 
should be replaced by $\sharp$.}  $M=\cV^\dagger \cV$ and $A=M^{-1} dM$ 
that (\ref{deltag}) implies
\be
\delta M = M\eta + \eta^\dagger M\,,\qquad \delta A= D\eta + 
            D(M^{-1} \eta^\dagger M)\,,\label{delMA}
\ee
where the $\bG$-covariant exterior derivative is defined on any $\cG$-valued
function $f$ by
\be
Df = df + [A,f]\,.
\ee
It can also be seen from the definition of $\eta$ in (\ref{deltag}), 
after making use of the Lax equation (\ref{Lax2}), that
\be
D\eta= \fft1{t}\, {*d\eta}\,,\qquad D(M^{-1} \eta^\dagger M)= 
            t\, {*d(M^{-1} \eta^\dagger M)}\,.
\ee
Thus we conclude that under (\ref{deltag}),
\be
\delta A = {*d}\Big( \fft1{t} \, \eta + t\, M^{-1} \eta^\dagger M\Big)\,,
\ee
which indeed verifies that $d{*\delta A}=0$.  

   In order to read off the symmetry algebra one needs to calculate commutators
of the form $[\delta_\m,\delta_\n]$.  Since, as noted above, the variations
$\delta_\n$ involve $X(t)$, which itself depends non-locally on the fields of
the scalar coset, one first needs to calculate the variations of $X(t)$ 
with respect
to the hierarchy of transformations $\delta_\n$.  This was obtained in
\cite{schwarz1}, and with a small but important modification that we shall
discuss later, it is given by
\be
\delta_1 X_2 = \fft{t_2}{t_1-t_2} \, (\eta_1 X_2 - X_2 \ep_1) +
   \fft{t_1 t_2}{1-t_1 t_2}\, M^{-1} \eta_1^\dagger M X_2\,.\label{del1X2}
\ee
Here $\delta_1$ (with no parentheses around the 1) denotes
$\delta(\ep_1,t_1)=\sum_{n\ge0} t_1^n \delta_\n(\ep_1)$, 
whilst $X_2$ denotes $X(t)$ for a different and independent
choice of spectral parameter $t_2$.  
    By equating the coefficients of $t_1^m t_2^n$
on both sides of (\ref{del1X2}), one can read off the variation under 
$\delta_\m$ of the $t_2^n$ term in the series expansion of $X(t_2)$.

   In order to derive (\ref{del1X2}), we follow the method used in 
\cite{schwarz1}, which amounts to varying the Lax equation (\ref{Lax2}) under
(\ref{deltag}), with $\delta A$ given in (\ref{delMA}) and $\delta X$ 
given by (\ref{del1X2}), and verifying that the varied equation is also
satisfied. Thus, one must substitute (\ref{del1X2}) into 
\be
[t_2(d+A) -{*d}](\delta_1 X_2) + t_2(\delta_1 A) X_2=0\,,
\ee
or in other words, after using (\ref{delMA}), into
\be
[t_2(d+A) -{*d}](\delta_1 X_2) + t_2 [ D\eta_1 + D(M^{-1}\eta_1^\dagger M)]
                     X_2=0\,.\label{varlax}
\ee
After some algebra, again involving the use of the Lax equation, the desired
result follows.

   Using (\ref{del1X2}) one can calculate the commutator of transformations
on $M=\cV^\dagger\, \cV$, finding (in a similar manner to \cite{schwarz1}) that
\be
[\delta_1,\delta_2] M= 
 \fft{t_1\, \delta(\ep_{12}, t_1)-t_2\, \delta(\ep_{12},t_2)}{t_1-t_2}\, M
\,,\label{comM}
\ee
where $\ep_{12} =[\ep_1,\ep_2]$.
It is also straightforward to show, after some lengthy algebra, that
\be
[\delta_1,\delta_2] X_3=
 \fft{t_1\, \delta(\ep_{12}, t_1)
   -t_2\, \delta(\ep_{12},t_2)}{t_1-t_2}\, X_3
\,,\label{comX}
\ee

   If the transformations $\delta$ given in (\ref{deltag}) and (\ref{del1X2}) 
were the only ones extending $G$ then we would have essentially ``half'' of
the affine Kac-Moody extension $\hat G$.  However, there are additional
transformations, which we shall denote by $\td\delta$, that also extend $G$.
These leave $M$ invariant but they do act non-trivially on $X$.  They are 
given by
\be
\td\delta_1 M=0\,,\qquad \td\delta_1 X_2 = \fft{t_1 t_2}{1-t_1 t_2} 
                  X_2 \ep_1\,.\label{tddeldef}
\ee
Again, the notation here is that $\td\delta_1 =\td\delta(\ep_1,t_1)
=\sum_{n\ge 1} t_1^n \td\delta_\n$, and $X_2=X(t_2)$.  (Note that
there is no $n=0$ term here in the expansion of $\td\delta_1$, as
can be seen from the absence of a $t_1^0$ term on the right-hand side of
(\ref{tddeldef}).)  It is easy to verify that (\ref{tddeldef}) describes
symmetries of the Lax equation.  The easiest way to do this is to
note that (\ref{tddeldef}) implies $\td\delta (dX X^{-1}) =0$, and
so since $\td\delta A=0$, it is evident that the Lax equation (\ref{Lax3})
is indeed invariant under $\td\delta$.

   The commutators of the $\td\delta$ transformations give
\be
[\td\delta_1,\td\delta_2] X_3 = \fft{t_2}{t_1-t_2}\, 
  \td\delta(\ep_{12},t_1) X_3
  -\fft{t_1}{t_1-t_2}\, \td\delta(\ep_{12}, t_2) X_3 
        \,,\label{tdtdcom}
\ee
where again, $\ep_{12}=[\ep_1,\ep_2]$.  (This commutation relation is vacuous,
of course, when acting on $M$.)   Finally, we may calculate
the commutators of $\delta$ and $\td\delta$ transformations, finding
\be
{[}\delta_1,\td\delta_2{]} X_3 = \fft{t_1 t_2}{1-t_1 t_2}\, 
               \delta(\ep_{12},t_1) X_3 + \fft1{1-t_1 t_2}\, 
          \td\delta(\ep_{12},t_2) X_3\,.\label{deltdcom}
\ee
(The commutator on $M$ is the same, except that there is no $\td\delta$
term on the right-hand side since $\td\delta M=0$.)

   In summary, we therefore have in total the commutation relations
\bea
{[}\delta_1,\delta_2{]} &=&
 \fft{t_1}{t_1-t_2}\, 
      \delta(\ep_{12}, t_1)- \fft{t_2}{t_1-t_2}\, 
   \delta(\ep_{12},t_2)\,,
\label{nn}\\
{[}\delta_1,\td\delta_2{]} &=& \fft{t_1 t_2}{1-t_1 t_2}\,
               \delta(\ep_{12},t_1) + \fft1{1-t_1 t_2}\,
          \td\delta(\ep_{12},t_2)\,,\label{ntd}\\
{[}\td\delta_1,\td\delta_2{]} &=& \fft{t_2}{t_1-t_2}\,
  \td\delta(\ep_{12},t_1)
  -\fft{t_1}{t_1-t_2}\, \td\delta(\ep_{12}, t_2)\,.\label{tdtd}
\eea
 From these, one can read off the towers of modes in the $t$-expansions,
using $\delta(\ep,t)=\sum_n t^n \delta_n(\ep)$, etc.  For example,
(\ref{nn}) gives
\bea
\sum_{m\ge0} \sum_{n\ge0} t_1^m t_2^n\, [\delta_\m(\ep_1),\delta_\n(\ep_2)]
  &=& \fft1{t_1-t_2}\, \sum_{p\ge0} (t_1^{p+1}-t_2^{p+1})\,
   \delta_\p(\ep_{12})\,,\nn\\
&=& \sum_{p\ge0} \sum_{q=0}^p t_1^q t_2^{p-q}\, 
        \delta_\p(\ep_{12})\,,\nn\\
&=& \sum_{m\ge0} \sum_{n\ge0} t_1^m t_2^n\, \delta_{\sst{(m+n)}}(\ep_{12})\,,
\eea
whence we obtain
\be
[\delta_\m(\ep_1),\delta_\n(\ep_2)] = \delta_{\sst{(m+n)}}(\ep_{12})\,,
\qquad m\ge0\,,\ n\ge0\,.\label{nn2}
\ee
The analogous calculations for (\ref{ntd}) and (\ref{tdtd}) give
\bea 
 [\delta_\m(\ep_1),\td\delta_\n(\ep_2)] &=& \delta_{\sst{(m-n)}}(\ep_{12})
        + \td\delta_{\sst{(n-m)}}(\ep_{12})\,,\qquad m\ge0\,,\ n\ge1\,,
   \label{ntd2}\\
 {[}\td\delta_\m(\ep_1),\td\delta_\n(\ep_2){]} &=&
       \td\delta_{\sst{(m+n)}}(\ep_{12})\,,
  \qquad m\ge1\,,\ n\ge1\,,\label{tdtd2}
\eea
where in (\ref{ntd2}) it is to be understood that $\delta_\n=0$ for
$n\le -1$ and $\td\delta_\n=0$ for $n\le0$.

    The three sets of commutation relations can be combined into one
by introducing a new set $\Delta_\n$ of variations, defined for all $n$
with $-\infty\le n\le \infty$, according to
\bea
   \Delta_\n &=& \delta_\n\,,\qquad n\ge0\,,\nn\\
 \Delta_{\sst{(-n)}} &=& \td\delta_\n\,,\qquad n\ge 1\,.\label{Deltadef}
\eea
It is then easily seen that (\ref{nn2}), (\ref{ntd2}) and (\ref{tdtd2})
become
\be
[\Delta_\m(\ep_1),\Delta_\n(\ep_2)] = \Delta_{\sst{(m+n)}}(\ep_{12})\,,
\qquad m,n\in\Z \,,\label{kacmoody}
\ee
with $\ep_{12}=[\ep_1,\ep_2]$.
This defines the affine Kac-Moody algebra $\hat \cG$.  In terms of currents
$J^i(\sigma)$ defined on a circle, with
\be
  J^i(\sigma) = \sum_{n=-\infty}^\infty e^{in\sigma}\, J_n^i\,,
\ee
the commutation relations (\ref{kacmoody}) are equivalent to
\be
 [J_m^i, J_n^j] = f^{ij}{}_k\, J^k_{m+n}\,,\label{kacmoody2}
\ee
where $f^{ij}{}_k$ are the structure constants of the Lie algebra $\cG$.
Specifically, we have the association
\be
\Delta_\n(\ep^i) \leftrightarrow J_n^i\,,\label{DeltaJ}
\ee
where $\ep= \ep^i\, T_i$, and $T_i$ are the generators of the Lie algebra
$\cG$.

   Since we have arrived at a somewhat different conclusion from Schwarz,
who finds only a subalgebra of the Kac-Moody algebra $\hat\cG$ as a 
symmetry of the SSM \cite{schwarz1}, we shall discuss in appendix \ref{appA} 
exactly why the difference has arisen.  In essence, the key distinction is
that we include the transformations $\td\delta$ defined in (\ref{tddeldef})
as independent symmetries.
They are non-trivial symmetries of the Lax equation, even though they
act trivially on the scalar fields in the coset representative $\cV$ itself.
In section \ref{sl2rsec}, we shall study the explicit
example of the $SL(2,\R)/O(2)$ 
coset model, in order to illustrate this point in greater detail.  We shall
show that a natural formulation of the model involves introducing an 
infinite number of additional scalar fields, in terms of which
$X$ appearing in the Lax equation (\ref{Lax2}) can be expressed as a
local quantity.  The $\td\delta$ transformations act on this infinite tower
of additional fields.  We shall also show how this infinity of extra scalars
can be interpreted as fields that one introduces in order to exhibit in a
local fashion the symmetries arising from
the closure of the two non-commuting $SL(2,\R)$ symmetries of the 
original theory and a dualised version.

   A further remark about the Kac-Moody transformations $\delta$ and 
$\td\delta$ is also in order.  The $\td\delta$ transformation defined
in (\ref{tddeldef}) are of the general form $\td\delta X \sim X\ep$.
It can be seen that the second of the three terms on the right-hand side
of the $\delta X$ transformation given in (\ref{del1X2}) is also of this
general form.  This means that as far as obtaining symmetries of the
Lax equation is concerned, one could have omitted the second term in
(\ref{del1X2}) altogether, since it is itself a distinct symmetry in
its own right.  However, it actually serves an important purpose in
(\ref{del1X2}), namely to subtract out what would otherwise be a pole
at $t_1=t_2$ if one had only $t_2 \eta_1 X_2/(t_1-t_2)$ rather than
$t_2(\eta_1 X_2 - X_2\ep_1)/(t_1-t_2)$.  (The third term is in (\ref{del1X2})
is necessary in addition, in order to get a symmetry, but there is no
pole associated with this term, since we expand $t_1$ and $t_2$ around zero.)
Now, the derivations of the $\delta X$ and $\td\delta X$ 
transformations as symmetries involved considering the variation of 
$(dX X^{-1})$ in the Lax equation (\ref{Lax3}).  In the case of
the $\td\delta$ transformation we have $\td\delta A=0$,
and one may view $\td\delta X$ as the solution of the homogeneous equation
$\td\delta(dX X^{-1})=0$, whilst $\delta X$ is the solution of the
inhomogeneous equation $\delta(dX X^{-1}) =(\hbox{non-zero source})$.  
Thus the inclusion of a $\td\delta X$
contribution as the second term in (\ref{del1X2}) can be viewed as the
necessary addition of a solution of the homogeneous
solution that is needed in order to ensure that the inhomogeneous 
solution satisfies the necessary boundary condition (\ie that 
$\delta_1 X_2$ be regular at $t_1=t_2$).  

   This discussion also emphasises
the point that it is really the $\delta$ transformations found by
Schwarz, appearing in our slightly modified form in (\ref{del1X2}), that
lie at the heart of the Kac-Moody symmetries of the symmetric-space
sigma models.  The $\td\delta$ transformations, although they are of
course equally necessary in order to obtain the complete Kac-Moody
symmetry, are somewhat secondary in nature since they are already
present within the construction of the $\delta$ transformations.

   It is also worth remarking that we have obtained the full Kac-Moody 
algebra as a symmetry of the SSM by means of
a purely perturbative analysis, which involved a 
small-$t$ expansion of $X(t)$ around $t=0$.
One may also consider instead a large-$t$ expansion of $X(t)$,
around $t=\infty$.  The
result is in fact equivalent.  This can be seen by letting 
$t=\td t^{-1}$, whereupon the Lax equation (\ref{Lax3}) becomes
\be
dX X^{-1} = - \fft{\td t}{1-\td t^2}\, {*A} - \fft{1}{1-\td t^2}\, A\,.
\ee
If we let $X=M^{-1} (\wtd X^{-1})^\dagger$, we arrive at a
Lax equation that is identical in form to the original expression 
(\ref{Lax3}), namely
\be
d \wtd X 
  \wtd X^{-1} =\fft{\td t}{1-\td t^2}\, {*A} +\fft{\td t^2}{1-\td t^2}\, A
\,,
\ee
showing that the large-$t$ expansion is equivalent to the small-$\td t$ 
expansion.
One would therefore reach identical conclusions had one performed a
large-$t$ expansion instead of a small-$t$ expansion.   It would be 
interesting to study the regime where the small-$t$ and large-$t$ expansions
overlap.  Although the Lax equation is regular in both regions,
it becomes singular at $t=\pm 1$.  Even if such a non-perturbative
analysis could be performed, we would not necessarily
expect to find a larger symmetry algebra than the full Kac-Moody algebra, 
which is already found in our perturbative approach.

\subsection{Virasoro-like symmetry}\label{Virasorosubsec}

   The symmetry discussed in section \ref{kacmoodysec} is an 
infinite-dimensional extension of the manifest $\cG$ symmetry of the 
$\bG/\bH$ symmetric-space sigma model.  As such, the transformation 
parameters $\ep$ in (\ref{deltag}) are themselves $\cG$ valued.
There is an additional infinite-dimensional symmetry of the SSM, with 
transformation parameters that are singlets under $\cG$, which turns out to 
be a subalgebra of the Virasoro algebra.  Our discussion here again begins
by using an approach that is very close to that of Schwarz \cite{schwarz1},
although with certain modifications and elaborations.   

   The transformations in question act on the coset representative $\cV$ as 
follows\footnote{We should really include an infinitesimal parameter
as a prefactor in the definition of $\xi$ in 
equation (\ref{vsym}).  However, since it is a singlet
it plays no significant r\^ole, and so it may be omitted without any 
risk of ambiguity.} \cite{schwarz1}:
\be
\delta^V(t) \cV = \cV \xi \,,\quad 
  \hbox{where}\quad 
\xi = -t \dot X(t) X(t)^{-1}\,.\label{vsym}
\ee
By equating the coefficients of each power of $t$ in
(\ref{vsym}), one obtains an infinite set of transformations $\delta^V_\n$
of the scalar fields in the SSM, with\footnote{Our transformations 
(\ref{vsym}) differ slightly from those given in \cite{schwarz1}, in which
the lowest-order term is subtracted out and the overall $t$-dependent
factor is different.  Our choice for the explicit $t$-dependent 
factor is made so that the algebra takes the simplest possible form. 
The subtraction was shown to be necessary
in the context of principal chiral models in \cite{schwarz1}, and was carried
over into the discussion of the SSM case in that paper.  In fact, the 
subtraction becomes
optional in the SSM case, which amounts to saying that the SSM has an
additional mode in the symmetry transformation.  We
shall discuss this in further in appendix \ref{Vapp}.} 
\be
\delta^V(t)= 
          \sum_{n\ge 1} t^n \delta^V_\n\,.\label{delVn}
\ee
Note that it is because of the explicit $t$ factor in the definition of 
$\xi$ in (\ref{vsym}) that the sum in (\ref{delVn}) does not include
$n=0$.

    To see that (\ref{vsym}) indeed describes symmetries of the theory, 
one must show that the equation of motion $d{*A}=0$ is preserved.  It 
follows from (\ref{vsym}) that
\be
\delta^V A= D\xi + M^{-1} d\xi^\dagger M=D\xi +
   D(M^{-1} \xi^\dagger M)\,,\label{delVA}
\ee
where as usual $D\xi = d\xi + [A,\xi]$.
Differentiating the Lax equation (\ref{Lax2}) with respect to $t$, and 
subtracting the Lax equation premultiplied by $(\dot X X^{-1})$ and 
postmultiplied by $X^{-1}$, one finds that
\be
D(\dot X X^{-1}) = \fft1{t} \,\Big[ {*d}(\dot X X^{-1}) -
       \fft1{1-t^2}\, A - \fft{t}{1-t^2}\,
                                        {*A}\Big]\,,\label{DXX}
\ee
 From this, one can also show that
\be
D(M^{-1} (\dot X X^{-1})^\dagger M) = 
  t\, {*d}(M^{-1} (\dot X X^{-1})^\dagger M) + \fft1{1-t^2}\, {*A} +
      \fft{t}{1-t^2}\, A\,.\label{DMXX}
\ee
Substituting into (\ref{delVA}), we find
\be
\delta^V A= {*d}\left( \fft1{t} \xi + t M^{-1} \xi^\dagger M\right)
         +  A\,,\label{delVA2}
\ee
and from this is follows that $d{*\delta^V A}=0$, thus proving that $\delta^V$
is a symmetry of the equations of motion.\footnote{It is because of the
cancellation in (\ref{delVA2}) of the contributions
proportional to ${*A}$ coming from the two terms in (\ref{delVA}) that there
is no need to make the lowest-order subtraction that was found in 
\cite{schwarz1} to be necessary in the PCM case.}

   The next step is to calculate the commutator of the $\delta^V$ 
transformations, in order to determine their algebra.  As a preliminary,
we need an expression for $\delta_1^V X_2$. 
Guided by the discussion in \cite{schwarz1}, we find that it is given by
\be
\delta_1^V X_2= Y X_2\,,\qquad
Y=\fft{1}{t_1-t_2}\, \Big[t_2\, \xi_1 + \fft{t_1(t_2^2-1)}{1-t_1 t_2}\, 
\xi_2\Big] 
  + \fft{t_1 t_2}{1-t_1 t_2}\, M^{-1} \xi_1^\dagger M
     \,.\label{delVX}
\ee
The verification
that (\ref{delVX}) is correct is
achieved by substituting (\ref{delVA}) and (\ref{delVX}) into the Lax 
equation (\ref{Lax3}).  

   After lengthy calculations of the commutators $[\delta_1^V,\delta_2^V]M$
and $[\delta_1^V,\delta_2^V] X_3$, we find that
\be
[\delta_1^V, \delta_2^V] =
-2 t_1 t_2 \left[ \fft{1}{(t_1-t_2)^2} +\fft1{(1-t_1 t_2)^2}\right]
  \, \delta_1^V 
  +\fft{t_1t_2(1-t_1^2)}{(t_1-t_2)(1-t_1 t_2)} \, \dot\delta_1^V 
-[1\leftrightarrow 2]\,,\label{comVM}
\ee
where $\dot\delta_1^V$ denotes the derivative of $\delta_1^V$ with respect
to its argument $t_1$, and the symbol $[1\leftrightarrow 2]$ indicates the
subtraction of two terms obtained from those that are displayed by 
exchanging the 1 and 2 subscripts everywhere.  

   To derive the mode algebra, we substitute the mode 
expansion (\ref{delVn}) into (\ref{comVM}), and
collect terms associated with each power of $t_1$ and $t_2$.  We then find
that the abstract algebra of the $\delta^V$ transformations is given by
\be
[\delta^V_\m, \delta^V_\n]= (m-n)\delta^V_{\sst{(m+n)}} 
              -    (m+n) \delta^V_{\sst{(m-n)}}\,,\label{Valg1}
\ee
where it is understood that $\delta^V_\n$ with negative mode numbers
$n$ is {\it defined} to be given by
\be
\delta^V_{\sst{(-n)}} \equiv -\delta^V_\n\,,\qquad n\ge1\,.\label{ref1}
\ee
One might have thought that the ostensible occurrence of
pole terms at $t_1=t_2$ in (\ref{comVM}) would have presented difficulties
in interpreting the algebra, but in fact one finds that cancellations 
imply there are no such
poles.  One way to make this manifest 
is to note that (\ref{comVM}) can be rewritten as
\be
[\delta_1^V,\delta_2^V] = -
 \fft{2t_1 t_2(\delta_1^V-\delta_2^V)}{(1-t_1 t_2)^2}
   -\fft{t_1 t_2(t_1-t_2)}{1-t_1 t_2}\,
  \fft{\del^2}{\del t_1\del t_2}\Big[\fft{t_2(t_1-t_1^{-1})\delta_1^V
   -t_1(t_2-t_2^{-1})\delta_2^V}{t_1-t_2}\Big]\,.
\ee

   We may define a current $K(\sigma)$ in which we associate 
the mode $K_n$ with the symmetry transformation $\delta^V_\n$:
\be
K(\sigma)=\sum_{n=-\infty}^\infty e^{in\sigma}\, K_n\,.\label{Kcurrent}
\ee
The reflection condition (\ref{ref1}) implies that the modes $K_n$ 
satisfy $K_n=-K_{-n}$, and from (\ref{Valg1}), they satisfy the algebra
\be
[K_m,K_n]= (m-n) K_{m+n} - (m+n) K_{m-n} \,,\label{Valg2}
\ee
This is clearly not the Virasoro algebra, but it is closely related to it.
Specifically, if we introduce generators $L_m$ for a centreless
Virasoro algebra,
\be
[L_m, L_n]= (m-n) L_{m+n}\,,
\ee
then we find that the modes $K_m$ may be represented as
\be
K_m= L_{m} - L_{-m}\,,\qquad m\ne 0\label{KL}
\ee
(Recall that that (\ref{vsym}) contains no $\delta^V_{(0)}$ transformation,
and so $K_0$ is not present in the algebra.)
If we define the usual Virasoro current
\be
T(\sigma)= \sum_{m=-\infty}^\infty L_m\, e^{im\sigma}\,,
\ee
then it follows from (\ref{Kcurrent}) and (\ref{KL}) that
\be
K(\sigma) =  2i\, \Im\left(T(\sigma)\right)\,.\label{KT2}
\ee

   It is interesting to contrast this result with the analogous one
that was obtained in \cite{schwarz1} for the case of a principal chiral model,
where it was shown that $K_m= L_{m+1} - L_{m-1}$ and hence
$K(\sigma)=-2i\sin\sigma\,
 T(\sigma)$.  In that case, one could view this relation as a definition of
the energy-momentum tensor $T(\sigma)$ in terms of $K(\sigma)$, save for the
degenerate points $\sigma=0$ and $\sigma=\pi$ at the ends of the line
segment.  By contrast, the relation (\ref{KT2}) for the SSM cannot be
used to define the whole of $T(\sigma)$, but only its imaginary part. 
Thus the Virasoro algebra itself is not described by the
symmetry transformations (\ref{vsym}). 

   We may also calculate the commutators of the Virasoro-like transformations
$\delta^V$ with the Kac-Moody transformations $\delta$ and $\td\delta$ of
section \ref{kacmoodysec}.  These commutators must be evaluated on 
$X$, and not merely on $M$, in order to capture the resulting 
terms that correspond to $\td\delta$ transformations, since $M$ is inert
under these.

   By calculating the commutator $[\delta_1^V, \td\delta_2]$ acting on $M$
and on $X_3$, we find that
\bea
[\delta_1^V, \td\delta_2] &=& \fft{t_1t_2 }{(1-t_1t_2)^2}\, 
\delta(t_1,\ep_2) +
  t_1t_2\left[\fft1{(1-t_1 t_2)^2} +\fft1{(t_1-t_2)^2}\right] \,
     \td\delta(t_2,\ep_2)\nn\\
&&- \fft{t_1t_2}{(t_1-t_2)^2}\, 
  \td\delta(t_1,\ep_2)  - 
            \fft{t_1t_2(t_2^2-1)}{(t_1-t_2)(1-t_1 t_2)}\, 
               \dot{\td\delta}(t_2,\ep_2)\,,\label{comm1}
\eea
where $\dot{\td\delta}(t_2,\ep_2)$ denotes the derivative of 
$\td\delta(t_2,\ep_2)$ with respect to $t_2$.

   Similarly, calculating the commutator $[\delta_1^V, \delta_2]$ acting
on $M$ and on $X_3$, we find 
\bea
[\delta_1^V, \delta_2] &=& \fft{t_1t_2}{(1-t_1t_2)^2}\,
\td\delta(t_1,\ep_2)  +
  t_1t_2\left[\fft1{(1-t_1 t_2)^2} +\fft1{(t_1-t_2)^2}\right] \,
     \delta(t_2,\ep_2) \nn\\
&&- \fft{t_1t_2}{(t_1-t_2)^2}\,
  \delta(t_1,\ep_2)  -
            \fft{t_1t_2(t_2^2-1)}{(t_1-t_2)(1-t_1 t_2)}\,
               \dot{\delta}(t_2,\ep_2) \,,\label{comm2}
\eea
As in the case of (\ref{comVM}), although there are ostensibly poles in
(\ref{comm1}) and (\ref{comm2}) at $t_1=t_2$, these in fact cancel.
Expanding in powers of $t_1$ and $t_2$, and making use of the definition
(\ref{Deltadef}) for the full set of Kac-Moody transformations $\Delta_m$,
we find that
\be
[\delta^V_\m,\Delta_\n]= -n( \Delta_{\sst{(n+m)}} -
   \Delta_{\sst{(n-m)}} )\,.
\ee
In terms of the Kac-Moody current-algebra modes $J_n^i$ and Virasoro-like
modes $K_n$ that we introduced earlier, we therefore find   
\be
[K_m,J^i_n] = -n( J^i_{n+m} - J^i_{n-m})\,.
\ee
One may verify that this is consistent with the Jacobi identity 
$[K_m,[K_n,J^i_p]]+\cdots=0$, after using our result (\ref{Valg2}) for
the commutator $[K_m,K_n]$.

\section{An Alternative Description}\label{altsec}

   A slightly different approach to describing the symmetries of 
two-dimensional symmetric-space coset models was taken in 
\cite{nicolai}, and it is useful to summarise some salient aspects here,
since we shall make use of some of the formalism in section \ref{sl2rsec}.  
It is
again an approach where the SSM is viewed as an integrable system, and it
is essentially equivalent to the description in \cite{schwarz1} in 
terms of the Lax equation.  

   Starting from the coset representative $\cV$ that we introduced
previously, one may define
\be
d\cV \cV^{-1} = Q + P\,,\label{QPdef}
\ee
where $Q$ is the projection into the denominator algebra $\cH$ and 
$P$ is the projection into the coset algebra $\cK$.  From the 
Cartan-Maurer equation $d(d\cV \cV^{-1})=(d\cV \cV^{-1})\wedge  
(d\cV \cV^{-1})$, one can then read off the equations
\bea
dQ - Q\wedge Q - P\wedge P &=&0\,,\label{CMH}\\
DP\equiv dP - Q\wedge P - P\wedge Q &=& 0\,.\label{CMK}
\eea
Under the transformations (\ref{gtrans}) one has
\be
Q\longrightarrow h Q h^{-1} + dh h^{-1}\,,\qquad
P\longrightarrow h P h^{-1}\,,\label{QPh}
\ee
which shows that $D=d-Q\wedge -\wedge Q$ can be viewed as an 
$\cH$-covariant connection. $P$ transforms covariantly under $\cH$ and
is invariant under the global right-acting $\cG$ transformations.

  From (\ref{MAdef}), and making the convenient assumption again that
the involution $\sharp$ is implemented by Hermitean conjugation, we
see that with $M=\cV^\dagger \cV$
\be
A=M^{-1} dM=
 \cV^{-1}\left(d\cV \cV^{-1} + (d\cV\cV^{-1})^\dagger \right)\cV =
\cV^{-1}(Q+P+ Q^\dagger + P^\dagger)\cV = 2\cV^{-1} P\cV\,,\label{AP}
\ee
since under the involution we shall have $Q^\dagger=-Q$, $P^\dagger=P$.
It follows from (\ref{Lax3}) that
\bea
\cV dX X^{-1} \cV^{-1} &=&
   \fft{2t}{1-t^2} \, {*P} + \fft{2t^2}{1-t^2} P\,,\nn\\
&=&  \fft{2t}{1-t^2} \, {*P} + \fft{1+t^2}{1-t^2} P -P\,,\nn\\
&=& \fft{2t}{1-t^2} \, {*P} + \fft{1+t^2}{1-t^2} P + Q -d\cV \cV^{-1}
 \,,
\eea
and hence
\be
d\hat\cV(t)  \hat\cV(t)^{-1} = Q +  \fft{2t}{1-t^2}\, {*P} +
                    \fft{1+t^2}{1-t^2} P\,, \label{QPP}
\ee
where we define
\be
\hat\cV(t)\equiv \cV X(t)\,.\label{VX}
\ee

   The Kac-Moody transformations $\delta$ and $\td\delta$, which we defined
in (\ref{deltag}), (\ref{del1X2}) and (\ref{tddeldef}), can now be
applied to $\hat\cV$.  We find
\bea
\delta_1 \hat\cV_2 &=& \fft{t_1}{t_1-t_2} \hat\cV_2 X_2^{-1} \eta_1 X_2 - 
           \fft{t_2}{t_1-t_2} \hat\cV_2 \ep_1 + 
  \fft{t_1 t_2}{1-t_1 t_2}\hat\cV_2 (M X_2)^{-1} \eta_1^\dagger M X_2
         + \delta h \hat \cV_2\,,\label{deltahatV}\\
\td\delta_1 \hat\cV_2 &=& \fft{t_1 t_2}{1-t_1 t_2} \hat\cV_2 \ep_1\,,
 \label{tddelhatV}
\eea
where as usual $\eta_1= X_1\ep_1 X_1^{-1}$, $\delta h$ is an $\cH$ 
compensating transformation and $\hat\cV_2=\cV X_2$.  

    The quantity $A=M^{-1} dM$ can be thought of as a $\cG$-valued 
conserved current, since as we noted in section (\ref{formsec}), it
transforms under global $\bG$ transformations $\cV\rightarrow h\cV g$ as
$A\rightarrow g^{-1} A g$, and it satisfies $d{*A}=0$.  We see from
(\ref{AP}) that $A=2\cV^{-1} P \cV$.  One can construct a hierarchy of  
conserved currents $\hat \cJ(t)$, for which $\hat \cJ(0)= A$, by defining
\be
\hat\cJ(t) =
 \fft{2}{1-t^2}\,\hat\cV^{-1} \left(\fft{1+t^2}{1-t^2} P + \fft{2t}{1-t^2}
        {*P}\right)\hat\cV\,.
\ee
That $\hat\cJ$ is conserved can be seen from the following calculation,
which also provides \cite{nicolai} a simpler expression for the currents:
\bea
\hat\cJ&=& \fft{2}{1-t^2}\,\hat\cV^{-1} *\left(\fft{2t}{1-t^2} P + 
        \fft{1+t^2}{1-t^2} {*P}\right)\hat\cV\,,\nn\\
&=& \hat\cV^{-1} * \fft{\del}{\del t}\left(\fft{1+t^2}{1-t^2} P
            + \fft{2t}{1-t^2} {*P} \right)\hat\cV\,,\nn\\
&=&\hat\cV^{-1} * \fft{\del}{\del t}\left(Q+ \fft{1+t^2}{1-t^2} P
            + \fft{2t}{1-t^2} {*P} \right)\hat\cV\,,\nn\\
&=&\hat\cV^{-1} * \fft{\del}{\del t}\left(d\hat\cV \hat\cV^{-1}\right) 
              \hat\cV\,,\nn\\
&=& {*d}\left(\hat\cV^{-1} \fft{\del\hat\cV}{\del t}\right)\,.
\label{Jdef}
\eea
Note that using (\ref{VX}), we can also write $\hat\cJ$ as
\be
\hat\cJ = {*d} (X^{-1} \dot X)\,.\label{Jdef2}
\ee
It is also useful to define the quantity
\be
{v}(t) = X^{-1}(t) \dot X(t) = \sum_{n\ge 0} t^n v_\n\,,\label{cvdef}
\ee
such that ${\cal J} ={*d}{v}$.

   The quantity ${v}(t)$ has a simple transformation under 
the $\td\delta$ Kac-Moody symmetries, with
\be
\td\delta_1 {v}(t_2) = \fft{t_1}{(1-t_1 t_2)^2}\, \ep_1  + 
      \fft{t_1 t_2}{1-t_1 t_2}\, [{v}(t_2),\ep_1] \,.\label{vtrans1}
\ee
In terms of the mode expansion in (\ref{cvdef}), this implies
\be
\td\delta_\m(\ep) {v}_\n = m \delta_{m,n+1} \ep +
               [{v}_{\sst{(n-m)}},\ep]\,.\label{vtrans2}
\ee

   The generalised currents $\hat\cJ={*d}{v}$ also transform 
nicely under the
Kac-Moody transformations $\td\delta$.  From (\ref{vtrans1}) we find
\be
\td\delta_1 \hat\cJ_2 = \fft{t_1 t_2}{1-t_1 t_2}\, [\hat\cJ_2,\ep_1]\,,
\label{tddelhatJ}
\ee
where $\hat\cJ_2 \equiv {*d}(X_2^{-1} \dot X_2)$.  If we expand $\hat\cJ$
as a power series
\be
\hat\cJ(t) = \sum_{n\ge0} t^n \hat\cJ_\n\,,
\ee
then (\ref{tddelhatJ}) implies that
\be
\td\delta_\m(\ep) \hat\cJ_\n = [\hat\cJ_{\sst{(n-m)}},\ep]\,,\qquad
  n\ge m\,.\label{Jtrans}
\ee
One might be tempted therefore to regard $\hat\cJ$ as defining a hierarchy of
Kac-Moody currents.  However, although they transform covariantly under
the ``lower half'' of the Kac-Moody symmetries corresponding to $\td\delta$,
their transformations in general under the ``upper half'' of the Kac-Moody
symmetries, corresponding to $\delta$, are very complicated, and one
cannot express $\delta_\m(\ep) \hat\cJ_\n$ as any linear combination of
$\hat\cJ_\p$ currents with field-independent coefficients.

\section{An Explicit Example: $SL(2,\R)/O(2)$ Coset Model}\label{sl2rsec}

\subsection{Infinitely many fields} 

  The simplest non-trivial example that illustrates the constructions
we have described in this paper is provided by the symmetric-space sigma
model $SL(2,\R)/O(2)$.  We begin by defining the $SL(2,\R)$ generators
\be
H=\begin{pmatrix} 1 & 0\cr
           0 & -1 \end{pmatrix}\,,\qquad
E^+=\begin{pmatrix}0& 1\cr
             0 & 0\end{pmatrix}\,,\qquad
E^-=\begin{pmatrix}0 & 0\cr
            1 & 0\end{pmatrix}\,.\label{sl2rgen}
\ee
The $O(2)$ denominator group is generated by the anti-Hermitean combination
$E^+ -E^-$, whilst the generators in the coset are the Hermitean matrices
$H$ and $E^+ +E^-$.  A convenient way to parametrise the coset representative
$\cV$ is in the Borel gauge, for which
\be
{\cal V}= e^{\ft12 \phi_0 H} e^{\chi_0 E^+}\,.\label{Vdef}
\ee
The fields $\phi_0$ and $\chi_0$ are the standard dilaton and axion of the
$SL(2,\R)/O(2)$ sigma model, with the Lagrangian
\be
L=-\ft14 \tr(A^\mu A_\mu) = -\ft12(\del\phi_0)^2 -\ft12 e^{2\phi_0}
             (\del\chi_0)^2\,.\label{sl2rlag}
\ee
 From (\ref{QPdef}) we find
\be
Q= \ft12(E^+-E^-) \wtd Q\,,\qquad P= \ft12 H P_\phi + \ft12(E^++E^-) P_\chi\,,
\ee
with
\be
\wtd Q = e^{\phi_0} d\chi_0\,,\qquad P_\phi= d\phi_0\,,\qquad
        P_\chi= e^{\phi_0} d\chi_0\,.\label{QPres}
\ee

   The standard $SL(2,\R)$ symmetry of the sigma model is given by
\be
\delta(\ep) \cV= \delta h \cV + \cV \ep\,,\label{sl2rsym0}
\ee
with $\ep= \ep^0 H +
  \ep^- E^+ + \ep^+ E^-$, where $\delta h$ is the appropriate $O(2)$
compensating transformation to restore the Borel gauge choice.  Thus
we have
\be
\delta\phi_0 = 2 \ep^0 + 2\ep^+ \,\chi_0\,,\qquad
  \delta\chi_0 = \ep^- - 2\ep^0\,\chi_0  + \ep^+ \, (e^{-2\phi_0} -\chi_0^2)\,.
\label{sl2rsym1}
\ee

  The next step is to define $\hat\cV$, whose relation to $X$ is given in
(\ref{VX}).  Following the general idea described in \cite{nicolai}, we
do this by introducing scalar fields $\hat\phi$, $\hat\chi$ and $\hat\psi$,
which depend on the spectral parameter $t$ as well as the spacetime
coordinates, and writing
\be
\hat\cV(t) =  e^{\ft12\phi(t) H} e^{\chi(t) E^+}
  e^{\psi(t) E^-} \,.\label{hatVdef}
\ee
We require that $\hat\cV$ smoothly approach $\cV$, defined in (\ref{Vdef}),
as $t$ goes to zero, and so
\be
\phi(0)=\phi_0\,,\qquad \chi(0)=\chi_0\,,\qquad \psi(0)=0\,.
\ee
In terms of power-series expansions for $\phi$, $\chi$ and $\psi$,
we may therefore write
\bea
\phi(t) &=& \phi_0 + t \phi_1 + t^2 \phi_2 + \cdots\,,\nn\\
\chi(t) &=& \chi_0 + t \chi_1 + t^2 \chi_2 +\cdots \,,\nn\\
\psi(t) &=& t \psi_1 + t^2 \psi_2 + \cdots\,.\label{fieldexp}
\eea

   Since $Q$, $P$ and ${*P}$ in (\ref{QPP}) are independent of the spectral
parameter $t$, it follows that by substituting (\ref{hatVdef}) into
(\ref{QPP}) we can read off a hierarchy of equations for the fields
$\phi_i$, $\chi_i$ and $\psi_i$.  At order $t^0$, we simply obtain the
expressions for $\wtd Q$, $P_\phi$ and $P_\chi$ already given in (\ref{QPres}).
At order $t^1$, we find
\bea
{*P_\phi} &=& \ft12 d\phi_1 + \chi_0 d\psi_1\,,\label{alev1phi}\\
{*P_\chi} &=& \ft12 e^{\phi_0}( d\chi_1 + \phi_1 d\chi_0 -
                 \chi_0^2 d\psi_1) + \ft12 e^{-\phi_0} d\psi_1\,,\\
0 &=& e^{\phi_0}( d\chi_1 + \phi_1 d\chi_0 -
                 \chi_0^2 d\psi_1) - e^{-\phi_0} d\psi_1\,,\label{alev1Q}
\eea
where the last equation comes from the absence of $t$-dependence in the
denominator group term $\wtd Q$.  It can be used to simplify the ${*P_\chi}$
expression, to give
\be
{*P_{\chi}}= e^{-\phi_0} d\psi_1\,.\label{alev1chi}
\ee
By equating the $t^1$ expressions (\ref{alev1phi}) and 
 (\ref{alev1chi}) for ${*P_\phi}$ and ${*P_\chi}$ to the duals of the $t^0$
expressions for $P_\phi$ and $P_\chi$ in (\ref{QPres}), we obtain, together
with (\ref{alev1Q}), the $t^1$ equations of motion
\bea
{*d\phi_0} &=& \ft12 d\phi_1 + \chi_0\, d\psi_1\,,\label{lev1phi}\\
e^{2\phi_0}\, {*d\chi_0} &=& d\psi_1\,,\label{lev1chi}\\
0 &=& d\chi_1 + \phi_1 d\chi_0 - (\chi_0^2 + e^{-2\phi_0}) d\psi_1\,.
  \label{lev1Q}
\eea

  At order $t^2$ we find
\bea
P_\phi &=& \ft12 d\phi_2 + \chi_1 d\psi_1 + \chi_0 d\psi_2\,,\label{lev2phi}\\
P_\chi&=& e^{-\phi_0}\, (d\psi_2 - \phi_1 d\psi_1)\,,\label{lev2chi}\\
0&=& d\chi_2 + \phi_1 d\chi_1 + (\phi_2+\ft12 \phi_1^2)d\chi_0
 -(\chi_0^2+e^{-2\phi_0}) d\psi_2 \nn\\
&&+[\phi_1 e^{-2\phi_0} -
     \chi_0(2\chi_1 + \phi_1 \chi_0)] d\psi_1\,,\label{lev2Q}
\eea
and we therefore obtain in total 3 equations at this order, after equating
these expressions for $P_\phi$ and $P_\chi$ to those in (\ref{QPres}).
One can continue this process to any desired order in $t$.

   The $SL(2,\R)$ symmetry $\delta(\ep)$ in (\ref{sl2rsym1}) extends to 
the higher-level
fields via the construction (\ref{del1X2}), with $t_1=0$.  Thus we have
$\delta(\ep) X= [X,\ep]$, and so using (\ref{VX}) to write 
$X=\cV^{-1} \hat\cV$, together with (\ref{hatVdef}), we find we can write the
$SL(2,R)$ transformations as
\bea
\delta \phi&=& -2 \ep^- \psi + 2\ep^0 \phi + 2\ep^+ \chi e^{\phi-\phi_0}\,,
\nn\\
\delta\chi &=& \ep^- (1+2\chi\psi) -2\ep^0 \chi + \ep^+e^{-\phi-\phi_0}
         (1-\chi^2 e^{2\phi})\,, \label{sl2rall}\\
\delta\psi &=& -\ep^- \psi^2 + 2\ep^0 \psi  + \ep^+ (1-e^{\phi-\phi_0})\,.\nn
\eea
Note that these transformations are linear when acting on $v$ defined in
(\ref{cvdef}): $\delta v= [v,\ep]$.

   Expanding out (\ref{sl2rall}) in powers of $t$, 
using (\ref{fieldexp}), we recover (\ref{sl2rsym1}) at order $t^0$, and 
at the next couple of orders we find
\bea
\delta \psi_1 &=& -\epsilon^+ \phi_1 + 2 \epsilon^0 \psi_1\,,\nn\\
\delta \phi_1&=&2\epsilon^+(\chi_1 + \chi_0\, \phi_1) -
2\epsilon^-\psi_1\,,\nn\\
\delta \chi_1 &=& -\epsilon^+ (2\chi_0\chi_1 +\chi_0^2 \phi_1 +
e^{-2\phi_0}\phi_1) - 2\epsilon^0 \chi_1 +2\epsilon^- \chi_0 \psi_1
\,,\nn\\
\delta \psi_2 &=& -\epsilon^+ (\phi_2 + \ft12\phi_1^2) +
2\epsilon^0\psi_2- \epsilon^- \psi_1^2\,,\nn\\
\delta \phi_2 &=& \epsilon^+ \Big(2\chi_2 + 2\chi_1\phi_1 +
\chi_0(\phi_1^2 + 2\phi_2)\Big) -2\epsilon^- \psi_2\,,\nn\\
\delta \chi_2 &=& \epsilon^+ \Big(-2\chi_0 (\chi_2 + \chi_1\phi_1) -
\ft12 \chi_0^2(\phi_1^2 + 2\phi_2) - \chi_1^2 +
e^{-2\phi_0} (\ft12 \phi_1^2 - \phi_2)\Big)\nn\\
&&\qquad - 2\epsilon^0 \chi_2 +
2\epsilon^- (\chi_1\psi_1 + \chi_0 \psi_2)\,.\label{sl2r3}
\eea
The hierarchy of equations of motion for the higher-level fields, for
which we presented the first two orders in (\ref{lev1phi})--(\ref{lev1Q}),
and (\ref{lev2phi})--(\ref{lev2Q}), are invariant under
the $SL(2,\R)$ transformations (\ref{sl2rall}).

\subsection{The Geroch group}\label{gerochsec}

   An interpretation of the higher-level fields can be given as follows.
The equations of motion for the original level-0 fields, following from
the Lagrangian (\ref{sl2rlag}), are
\be
d{*d\phi_0} + e^{2\phi_0}\, {*d\chi_0}\wedge d\chi =0\,,\qquad
  d(e^{2\phi_0} \, {*d\chi_0})=0\,,
\ee
Since we are in two dimensions, the axion $\chi_0$ can be dualised to 
another axion $\bar\chi_0$, such that
\be
d\bar\chi_0= e^{2\phi_0} \, {*d\chi_0}\,.\label{chidual}
\ee
Substituting this into the $\phi_0$ equation of motion, we can remove
a derivative from this equation too, obtaining
\be
{*d}\phi_0= d\sigma + \chi_0 d\bar\chi_0\,,\label{phidual}
\ee
for some new field $\sigma$.
Defining $\bar\phi_0=-\phi_0$, the original Lagrangian (\ref{sl2rlag}) can 
be written in a dualised form, terms of the barred fields, as
\be
{\cal L} = -\ft12(\del\bar\phi_0)^2 - \ft12 e^{2\bar\phi_0} (\del\bar\chi_0)^2
\,.\label{sl2rduallag}
\ee
We see, comparing (\ref{chidual}) and (\ref{phidual}) with (\ref{lev1phi})
and (\ref{lev1chi}), that 
\be
\bar \chi_0= \psi_1\,,\qquad \sigma= \ft12\phi_1\,.
\ee

   The dualised Lagrangian (\ref{sl2rduallag}) clearly also has an 
$SL(2,\R)$ symmetry, which we shall denote by $\overline{SL(2,\R)}$.  Denoting
its infinitesimal parameters by $\bep^\pm$ and $\bep^0$, this symmetry
acts on $\bar\phi_0$ and $\bar\chi_0$ exactly analogously to the action
of the original $SL(2,\R)$ on $\phi_0$ and $\chi_0$:
\be
\bar\delta\bar\phi_0 = 2 \bep^0 + 2\bep^- \,\bar\chi_0\,,\qquad
  \bar\delta\bar\chi_0 = \bep^+ - 2\bep^0\,\bar\chi_0  
       + \bep^- \, (e^{-2\bar\phi_0} -\bar\chi_0^2)\,.
\label{sl2rsymdual1}
\ee
(For notational reasons that will become clear shortly, we switch the 
$+$ and $-$ indices on $\bep^\pm$, relative to $\ep^\pm$, when passing to
this barred version of (\ref{sl2rsym1}).)  
  
   One may also define an infinite
tower of higher-level barred fields for the dualised sigma model, precisely  
analogous to the unbarred ones defined above.  For example, in order to 
obtain the
barred version of
(\ref{lev1phi})--(\ref{lev1Q}), we should make the 
identifications
\be
\bar\phi_0=-\phi_0\,,\qquad \bar\chi_0= \psi_1\,,\qquad \bar\phi_1 =
    -\phi_1 - 2\chi_0 \psi_1\,.\label{tilden}
\ee
The barring operation is an involution, with the bar of a bar being the
identity operator, and so there is an analogous version of (\ref{tilden})
in which all barred and unbarred fields are exchanged. The relations 
(\ref{tilden}) can be extended to all levels, as we shall now discuss.
 
   What we are seeing here is that although the original $(\phi_0,\chi_0)$
fields are non-locally related to the dual fields $(\bar\phi_0,\bar\chi_0)$
(because of the differential relation (\ref{chidual}) expressing $\bar\chi_0$
in terms of $\chi_0$), there exists a purely local relation between the
full hierarchy of fields $(\phi_i,\chi_i,\psi_i)$ and their barred 
analogues.  This relation can be established to any desired higher order in 
level 
number, by systematically examining the systems of equations that
follow from (\ref{QPP}), which we presented at level-1 in
(\ref{lev1phi})--(\ref{lev1Q}) and level-2 in  (\ref{lev2phi})--(\ref{lev2Q}).
There is, however, a simpler way of presenting the entire hierarchy
of relations in a compact form.

   To do this, we first introduce a barred version of $\hat\cV$, which was 
defined in equation (\ref{hattdVdef}):
\be
\hat{\bar\cV}(t) =  e^{\ft12 \bar\phi(t) \bar H} 
      e^{\bar\chi(t) \bar E^+}
  e^{\bar\psi(t) \bar E^-} \,.\label{hattdVdef}
\ee
Here $\bar H$ and $\bar E^\pm$ are $SL(2,\R)$ generators that satisfy
identical commutation relations to $H$ and $E^\pm$, namely
\be
[H,E^\pm]= \pm2 E^\pm\,,\quad [E^+,E^-]=H\,;\qquad
[\bar H,\bar E^\pm]= \pm2 \bar E^\pm\,,\quad [\bar E^+,\bar E^-]=\bar H\,.
\label{2copies}
\ee
This is already enough to ensure that the barred hierarchy of fields
will satisfy identical equations of motion to the unbarred hierarchy; 
they are derived
from the barred version of (\ref{QPP}).  Next, we note that we may make
the following choice for the barred generators in terms of the unbarred
ones:
\be
\bar E^+ = t\, E^-\,,\qquad \bar E^-= \fft1{t}\, E^+\,,\qquad
  \bar H=-H\,,
\ee
since this is consistent with (\ref{2copies}).  Thus we have
\be
\hat{\bar\cV}(t) =  e^{-\ft12\bar\phi(t) H}
      e^{t\, \bar\chi(t) E^-}
  e^{t^{-1}\, \bar\psi(t) E^+} \,.\label{hattdV2}
\ee

   We now impose the relation
\be
  \hat{\bar\cV}(t)= \hat\cV(t)\,
\ee
which therefore establishes a relation between these barred and unbarred
fields, which have already been established to satisfy the same system 
of equations.  This is easy to solve explicitly, since one has only
to exponentiate $2\times 2$ matrices in this example.  We find (suppressing
the explicit indication of the $t$-dependence of all the fields)
\be
\psi = 
 \fft{t \bar\chi}{1+\bar\chi\bar\psi}\,,\quad
\chi= \fft1{t}\, \bar\psi \left(1+\bar\chi\bar\psi\right)
\,,\quad
\phi=-\bar\phi - 
    2 \log\left(1+\bar\chi\bar\psi\right)\,.\label{allorders}
\ee

  Expanding in powers of $t$ allows us to read off the relation between
the entire hierarchies of barred and unbarred fields.  At the leading order,
we find precisely the relations (\ref{tilden}) that we obtained previously
when we started the level-by-level process of mapping the unbarred equations of
motion into barred ones.  If one carries out such a sequential calculation,
one finds that the entire hierarchy of relations between barred and
unbarred fields uniquely follows, once the leading-order relations 
(\ref{tilden}) are fed in.  Thus, we may conclude that since the
all-level relations (\ref{allorders}) match (\ref{tilden}) at the leading
order, they represent the unique completion of this relation to all orders.

   The barred hierarchy of fields transforms under $\overline{SL(2,\R)}$ in
precisely the same way as the unbarred hierarchy transforms under
$SL(2,\R)$.  For example, for the first couple of levels,
the barred fields will transform under the dual $\overline{SL(2,\R)}$ 
symmetry according to
the barred version of (\ref{sl2r3}) (with the exchange of $\bep^+$ and
$\bep^-$, as we discussed previously for $\bar\phi_0$ and $\bar\chi_0$ in
(\ref{sl2rsymdual1})).  The $\overline{SL(2,\R)}$ transformations of
the entire hierarchy of dual fields can be succinctly expressed as the
barred analogue of (\ref{sl2rall}), which is therefore given by
\bea
\bar\delta \bar\phi&=& 
  -2 \bep^+ \bar\psi + 2\bep^0 \bar\phi + 2\bep^- \bar\chi 
        e^{\bar\phi-\bar\phi_0}\,,
\nn\\
\bar\delta\bar\chi &=& \bep^+ (1+2\bar\chi\bar\psi) 
   -2\bep^0 \bar\chi + \bep^- e^{-\bar\phi-\bar\phi_0}
         (1-\bar\chi^2 e^{2\bar\phi})\,, \label{barsl2rall}\\
\bar\delta\bar\psi &=& -\bep^+ \bar\psi^2 
  + 2\bep^0 \bar\psi  + \bep^- (1-e^{\bar\phi-\bar\phi_0})\,.\nn
\eea

   Since we also have the relation (\ref{allorders}) between the barred and
the unbarred fields, it is now a straightforward matter to work out the
transformations of the original unbarred fields under the dual 
$\overline{SL(2,R)}$ symmetry.   From (\ref{allorders}) and 
(\ref{barsl2rall}) we find
\bea
\bar\delta \phi &=& -2\bep^0 - 2 \bep^- \left[ t \chi e^{\phi+\phi_0} 
      + \fft1{t} \psi\right ]\,,\nn\\
\bar\delta\chi &=& 2\bep^0 \chi  +\bep^- \left[ t\chi^2 e^{\phi+\phi_0}
       +\fft1{t} (1+2\chi\psi - e^{-\phi+\phi_0})\right]\,,\label{delbar}\\
\bar\delta\psi &=& t \bep^+ -2\bep^0 \psi +
   \bep^- \left[ t e^{\phi+\phi_0} -\fft1{t} \psi^2\right]\,.\nn
\eea
Expanded, as usual, in powers of $t$, these equations give the transformations
of the entire hierarchy of original fields $(\phi_i,\chi_i,\psi_i)$ under
the dual $\overline{SL(2,R)}$ symmetry.  Note that there are no negative
powers of $t$ in the expansions.

 It is evident from (\ref{delbar}) that the $\bep^0$ transformation in
$\overline{SL(2,\R)}$ is
the same (modulo a sign) as the $\ep^0$ transformation with respect to
the original $SL(2,\R)$ (see equation (\ref{sl2rall})).  The $\bep^+$
transformation in (\ref{delbar}) is also very simple, with
\be
\bar\delta(\bep^+)\phi=0\,,\qquad 
\bar\delta(\bep^+)\chi=0\,,\qquad
\bar\delta(\bep^+)\psi= t \bep^+\,.
\ee
In terms of the expansions (\ref{fieldexp}), this means that all fields
$(\phi_i,\chi_i,\psi_i)$ in the hierarchy are inert except for $\psi_1$,
which suffers the shift transformation
\be
\bar\delta(\bep^+) \psi_1 = \bep^+ \,.\label{bepp}
\ee
It is easy to see that this is precisely the same as the transformation
given by $\td\delta_1 X_2$ in equation (\ref{tddeldef}), at order $t_1^1$
and with $\ep_1$ taken to be just $\ep^+$, \ie
\be
\td\delta_{\sst{(1)}}(\bep^+) X_2 = t_2\, X_2\bep^+\,.
\ee
This shows that the $\bep^+$ transformation in $\overline{SL(2,\R)}$
is implemented by the Kac-Moody generator $J^+_{-1}$ (see (\ref{DeltaJ})).  

   This leaves the $\bep^-$ transformation in $\overline{SL(2,\R)}$
still to be identified.  In fact, this is precisely
a $\delta_1 X_2$ transformation as given in (\ref{del1X2}), at order $t_1^1$
and with $\ep_1$ taken to be just $\bep^-$.  Using (\ref{del1X2}), this
is given by 
\be
\delta_{\sst{(1)}}(\ep_1) X_2= \fft1{t_2} [X_2,\ep_1] - 
      \dot\eta_1(\ep_1,0) X_2 + t_2 M^{-1}\ep_1^\dagger M X_2\,,
\ee
with $\ep_1=\bep^-$, where $\eta_1(\ep_1,t)\equiv X(t) \ep_1 X^{-1}(t)$.  
Substituting $X=\cV^{-1}\hat\cV$ into this, and using (\ref{hatVdef}),
one straightforwardly reproduces the $\bep^-$ transformation in
(\ref{delbar}).  This shows that the $\bep^-$ transformation in 
$\overline{SL(2,\R)}$ is implemented by the Kac-Moody generator $J_1^-$
(see (\ref{DeltaJ})).

   At this stage, we have arrived at a complete understanding of all
six transformations in the original and dual symmetry groups $SL(2,\R)$
and $\overline{SL(2,\R)}$.  The original $SL(2,\R)$ transformations
$\ep^\pm$ and $\ep^0$ 
of course correspond to the level-0 Kac-Moody generators $J_0^\pm$ and
$J_0^0$.  We have also shown that the dual $\overline{SL(2,\R)}$ 
transformations $\bep^+$, $\bep^-$ and $\bep^0$ correspond
to the Kac-Moody generators $J_{-1}^+$, $J_1^-$ and $J_0^0$:
\bea
SL(2,\R):&& (J_0^+, J_0^-, J_0^0)\,,\nn\\
&&\nn\\
\overline{SL(2,\R)}:&& (J_{-1}^+, J_1^{-}, J_0^0)\,.
\eea
It is indeed clear from the Kac-Moody algebra (\ref{kacmoody2}) that both these
triplets selected from the generators $J^i_n$ form $SL(2,\R)$ subalgebras.
It is also clear that the two triplets do not commute.  In fact, from
the two triplets one can fill out the entire Kac-Moody algebra, by taking
appropriate sequences of multiple commutators. 

   Thus we have shown in a very
explicit and precise way that the affine $\widehat{SL(2,\R)}$ 
Kac-Moody symmetry of
the two-dimensional $SL(2,\R)/O(2)$ symmetric-space sigma model is generated
by taking multiple commutators of the two $SL(2,\R)$ symmetries of the
original and the dualised formulations of the theory.

   It is interesting to note that the entire ``negative half'' of the
Kac-Moody symmetry (\ie $J^i_n$ with $n<0$), which can be generated
by multiple commutation of $J^+_{-1}$ with $J^i_n$ with $n\ge0$, emerges 
from the humble shift symmetry $\bar\delta\psi_1=\bep^+$ that
we obtained in (\ref{bepp}).  This emphasises the point, which we remarked on
earlier, that the negative half of the Kac-Moody algebra arises through 
symmetries that are realised only on the infinite tower of fields 
$(\phi_i,\chi_i,\psi_i)$ with $i>0$ that were introduced in order to
allow the symmetries of the sigma model to be expressed in a local,
as opposed to non-local, manner. (See appendix \ref{appA} for further 
discussion of this point.)

\subsection{Explicit formulae for $\td\delta$ and some example $\delta$
    transformations}

   It is not hard to work out the explicit form of all the $\td\delta$
transformations on the fields $(\phi_i,\chi_i,\psi_i)$.  From 
(\ref{tddeldef}), (\ref{VX}) and (\ref{hatVdef}) we find
\bea
\td\delta_1 \psi(t_2) &=& \fft{t_1 t_2}{1-t_1 t_2}\, 
       \left(\ep^+ + 2\ep^0 \psi(t_2)  -\ep^-\,\psi(t_2)^2\right)\,,\nn\\
\td\delta_1 \chi(t_2) &=& \fft{t_1 t_2}{1-t_1 t_2}\,\left( -2\ep^0\chi(t_2) +
            \ep^-\, (1+2\chi(t_2)\psi(t_2))\right)\,,\nn\\
\td\delta_1\phi(t_2) &=& \fft{2t_1 t_2}{1-t_1 t_2}\,(\ep^0-\ep^-\, 
\psi(t_2))\,.
\eea
Collecting the powers of $t_1$ and $t_2$, we find for $n\ge m\ge1$ that 
\bea
\td\delta_\m(\ep) \phi_n &=& 2\delta_{mn}\, \ep^0 - 2\ep^-\, 
\psi_{n-m}\,, \nn\\
\td\delta_\m(\ep)\chi_n &=& -2\delta_{mn}\, \ep^0\, \chi_n 
  +\delta_{mn}\,  \ep^- 
              + 2 \ep^-\, \sum_{p=0}^{n-m-1} \chi_p\, \psi_{n-m-p}\,,
\nn\\
\td\delta_m(\ep) \psi_n &=& \delta_{mn}\, \ep^+ 
    +2 \ep^0 \psi_{n-m} -\ep^-\, \sum_{p=1}^{n-m-1} 
   \psi_p \, \psi_{n-m-p}\,, \label{tddeltrans}
\eea
where it is understood that on the right-hand side $\chi_n=0$ for $n<0$
and $\psi_n=0$ for $n<1$.  Note that $\td\delta_\m \phi_n=0$, 
$\td\delta_\m \chi_n=0$ and $\td\delta_\m \psi_n=0$ whenever $m<n$.  
Of course we also have $\td\delta_\m \phi_0=0$, $\td\delta_\m\chi_0=0$.

   The symmetries $\td\delta$ in (\ref{tddeltrans}) are essentially 
just shift transformations of $\phi_n$, $\chi_n$ and $\psi_n$ by constant 
parameters $\ep^0$,
$\ep^-$ and $\ep^+$ (with independent sets of these $SL(2,\R)$ 
parameters at each of the negative Kac-Moody levels), 
with the extra terms being the necessary 
``dressings'' that ensure that the transformations leave the equations
of motion invariant.  In accordance with an observation we made previously, 
the $\td\delta$ transformations could therefore be used in order to 
``gauge fix'' 
the auxiliary fields (\ie $(\phi_i,\chi_i,\psi_i)$ for $i\ge1$ in this 
$SL(2,\R)/O(2)$ example) to any desired set of values at one chosen point
in spacetime.  Since the auxiliary fields also transform under the
$\delta$ symmetries, one could view the $\td\delta$ transformations, in such
a gauge-fixed situation, as compensating transformations that restored
the fields to these chosen values after having performed $\delta$ 
transformations.  This is effectively what happens in the 
construction of Schwarz's subalgebra of the full Kac-Moody algebra.

   As we observed in section \ref{altsec}, the $\td\delta$ transformations
become more elegant if  they are applied to the quantities $v_\n$ defined
in (\ref{cvdef}), for which we have (\ref{vtrans2}).  In fact $v(t)$ is easily 
calculated in terms of $\phi(t)$,
$\chi(t)$ and $\psi(t)$, giving
\be
v^-= \dot\chi + \chi \dot\phi\,,\quad
v^0=\ft12 \dot\phi + \psi\dot\chi +\chi\psi\dot\phi\,,\quad
v^+=\dot\psi -(1+\chi\psi)\psi\dot\phi -\psi^2\dot\chi\,.
\ee
Thus, as can be seen by expanding in powers of $t$, the $v^\pm_\n$ and
$v_\n^0$ are are just certain combinations of the $\phi_m$, $\chi_m$ and
$\psi_m$ fields,
\be
v^-_{\sst{(0)}}= \chi_1+\chi_0\phi_1\,,\qquad v^0_{\sst{(0)}} = \ft12\phi_1
\,,\qquad v^+_{\sst{(0)}} = \psi_1\,,\qquad \hbox{etc.}
\ee

    The $\delta$ symmetries in (\ref{deltag}) and (\ref{del1X2}) are
more non-trivial, but again they are completely local transformations of
the fields $(\phi_i,\chi_i,\psi_i)$, which can be read off explicitly
to any desired order of non-negative Kac-Moody level, and to any desired order
in the $t$-expansion of the fields.  For example, we find for the
$SL(2,\R)/O(2)$ example that at Kac-Moody level 1, the
transformations on $(\phi_0,\chi_0,\psi_1,\chi_1,\psi_1)$ are given by
\bea
\delta_{\sst{(1)}}(\ep)\phi_0 &=& 2\ep^+ \, \chi_1 + 4\ep^0\, \chi_0 \psi_1
  - 2\ep^-\, \psi_1\,,\nn\\
\delta_{\sst{(1)}}(\ep)\chi_0 &=& 
         -\ep^+\, (\phi_1 e^{-2\phi_0} + 2\chi_0 \chi_1 
                       +\chi_0^2 \phi_1) +
 \ep^-\, (\phi_1 + 2\chi_0 \psi_1)\nn\\
&&
  + 2\ep^0\,(\psi_1 e^{-2\phi_0} - \chi_1 - \chi_0\phi_1 - \chi_0^2 \psi_1)
 \,,\nn\\
\delta_{\sst{(1)}}(\ep)\phi_1 &=& \ep^+\, \big(2\chi_2 + 
  \chi_0(2+2\phi_2 -\phi_1^2) + 2\chi_0^3 e^{2\phi_0}\big) +
   2\ep^-\, (\psi_2 + \chi_0 e^{2\phi_0}) \nn\\
&&+ 2\ep^0\, (1+ 2 \chi_0^2 e^{2\phi_0} + 2 \chi_1\psi_1 +
         2\chi_0 \phi_1\psi_1)\,,\nn\\
\delta_{\sst{(1)}}(\ep)\chi_1 &=&\ep^+\, \big( (1+\phi_1^2) e^{-2\phi_0}
  -\chi_1^2 - 2\chi_0 \chi_2 + \chi_0^2 (\phi_1^2 -2\phi_2) -
   \chi_0^4 e^{2\phi_0} ) \nn\\
&&+
 \ep^-\, (\phi_2 + 2\chi_0 \psi_2 + 2 \chi_1 \psi_1 -\ft12 \phi_1^2 +
    \chi_0^2 e^{2\phi_0}) \nn\\
 && + \ep^0 \,\big(-2 \chi_2 + \chi_0(\phi_1^2 -2\phi_2 -4\chi_1\psi_1) 
    -2\chi_0^2 \phi_1\psi_1 
 -2\phi_1\psi_1 e^{-2\phi_0} - 2\chi_0^3 e^{2\phi_0}\big)\,,\nn\\
\delta_{\sst{(1)}}(\ep)\psi_1 &=& 
 -\ep^+\, (\phi_2 -\ft12\phi_1^2 + \chi_0^2 e^{2\phi_0}) +
       \ep^-\, (e^{2\phi_0} - \psi_1^2) \nn\\
&& + 2\ep^0\,(\psi_2 -\phi_1\psi_1  - \chi_0 e^{2\phi_0})\,.
\eea 

\section{Conclusions}

   In this paper, we have studied the global symmetries of flat
two-dimensional symmetric-space sigma models.  This can be viewed as
a preliminary to studying the somewhat more intricate problem of 
curved-space two-dimensional sigma models, which arise in the
toroidal compactification of supergravity theories.  Both the curved and the 
flat cases share the common feature that the global symmetries include
an infinite-dimensional extension of the manifest $\cG$ symmetry of the
$\bG/\bH$ sigma model.  

     There has been some controversy over the precise nature of the 
infinite-dimensional extension. Whilst most authors have
asserted that the symmetry is the affine Kac-Moody extension $\hat\cG$ of 
$\cG$, Schwarz \cite{schwarz1} found instead a certain subalgebra
$\hat\cG_H$ of the Kac-Moody algebra.  One of our goals in this paper
has been to resolve the discrepancies.

   In our work we made extensive use of Schwarz's results which have,
it seems for the first time, provided explicit expressions for the
key transformations that underlie the positive half of the 
Kac-Moody symmetry algebra.
By synthesising this with earlier work where the idea of introducing an
infinity of auxiliary fields in order to provide a local formulation was
developed, we have been able to construct a fully local description of the
entire Kac-Moody algebra of global symmetry transformations.

   We have also shown how the subalgebra found by Schwarz can be viewed as
a consequence of making a gauge choice, in which the values of the
complete set of fields are fixed to prescribed values at a chosen 
distinguished point in the two-dimensional spacetime.
   
   In order to make some of the ideas more concrete, we also studied a
simple explicit example, where the coset of the sigma model is taken to be 
$SL(2,\R)/O(2)$.  We showed how our present analysis could
be related to much earlier work by Geroch \cite{geroch}, in which the
infinite-dimensional symmetry was obtained by commuting $SL(2,\R)$ 
symmetry transformations of the original sigma model and its dual version.
In particular, we were able to exhibit the precise correspondence between
the two sets of $SL(2,\R)$ transformations and certain generators of the
Kac-Moody algebra.  This provides an explicit demonstration that the Geroch
algebra formed by taking commutators of the two $SL(2,\R)$ transformations
is the same as the Kac-Moody algebra $\widehat{SL(2,\R)}$.

\section*{Acknowledgements}

   We are very grateful to John Schwarz for discussions, and for
drawing our attention to references \cite{schwarz1,schwarz2}.  We thank
also Gary Gibbons, Herman Nicolai and Kelly Stelle for discussions.  
This research has
been generously supported by George Mitchell and the Mitchell Family 
Foundation.  The research of 
H.L. and C.N.P. is also supported in part by DOE grant DE-FG03-95ER40917.

\appendix

\section{Schwarz Algebra $\hat\cG_H$ Versus Kac-Moody
        Algebra $\hat\cG$}
\label{appA}

   In \cite{schwarz1}, the Lax equation (\ref{Lax3}) is solved for $X$
as a non-local function of the original sigma-model fields, by writing
\be
X(x;t) = {\cal P} \exp\Big[\int_{x_0}^x\Big(
   \fft{t}{1-t^2}\, {*A} + \fft{t^2}{1-t^2} \, A\Big)\Big]\,,\label{pathi}
\ee
where ${\cal P}$ denotes path ordering along the integration path, and
$x_0$ is an arbitrarily-chosen point. This is a significantly different
approach from the one we have followed, where $X$ is expressed locally in
terms of an infinity of auxiliary fields.

  Our transformation (\ref{del1X2}) for $\delta_1 X_2$ is not quite the 
same as the one given in Schwarz's discussion \cite{schwarz1}.  Let us
denote his expression by $\delta_1' X_2$; it is given by
\be
\delta_1' X_2 = \fft{t_2}{t_1-t_2} \, (\eta_1 X_2 - X_2 \ep_1) +
   \fft{t_1 t_2}{1-t_1 t_2}\, (M^{-1} \eta_1^\dagger M X_2 -
      X_2 M_0^{-1} \ep_1^\dagger M_0)\,,\label{2del1X2}
\ee
where $M_0=M(x_0)$, and $x_0$ is chosen as the lower limit of the 
integral expression (\ref{pathi}) for $X(t)$.  Thus the relation between 
$\delta_1'$ and our expression $\delta_1$ is
\be
\delta_1' = \delta_1 - \fft{t_1 t_2}{1-t_1 t_2}\, X_2 
 M_0^{-1} \ep_1^\dagger M_0 \,.\label{schwtran}
\ee
In \cite{schwarz1}, Schwarz calculates the commutator $[\delta_1',\delta_2']
  M$, finding
\be
[\delta_1',\delta_2'] M=
 \fft{t_1\, \delta'(\ep_{12}, t_1)-t_2\, \delta'(\ep_{12},t_2)}{t_1-t_2}\, M -
   \fft{t_1\, t_2}{1-t_1\, t_2} \,
              \left(\delta'(\ep'_{12},t_1) -\delta'(\ep'_{12},t_2)\right)M\,,
\label{2comM}
\ee
where
\be
\ep_{12}= [\ep_1,\ep_2]\,,\qquad \ep'_{12}=
 [M_0^{-1} \ep_1^\dagger M_0, \ep_2]\,.\label{yyy}
\ee
(In obtaining this result, one must hold $M_0$ fixed.)
The right-hand side of (\ref{2comM}) involves $\delta'$ transformations 
again, and so the algebra appears to be closing. 
However, Schwarz does not calculate $[\delta_1',\delta_2'] X_3$.  Let
us denote his result in (\ref{2comM}) as $[\delta_1',\delta_2'] M=
\delta^S M$.  After some algebra, we find that
\be
[\delta_1',\delta_2'] X_3 = \delta^S X_3 + \fft{t_1 t_3}{1-t_1 t_3}\,
        X_3(M_0^{-1} \ep_{12}^\dagger M_0 -\ep_{12}') +
  \fft{t_2 t_3}{1-t_2 t_3}\,
        X_3(M_0^{-1} \ep_{12}^\dagger M_0 +\ep_{21}')\,.\label{schwX3}
\ee
This shows that on $X_3$, the commutator of $\delta'$ transformations does not
close merely on $\delta'$, but instead it gives transformations of the
form $X_3 \td\ep$ as well, for certain $\td\ep$.  In fact, such transformations
are of the type $\td\delta$ that we introduced in (\ref{tddeldef}),
and (\ref{schwX3}) may be written abstractly as
\bea
[\delta_1', \delta_2']=\delta^S + \td\delta(M_0^{-1}\epsilon_{12}^\dagger
M_0 - \epsilon_{12}', t_1) +
\td\delta(M_0^{-1}\epsilon_{12}^\dagger
M_0 + \epsilon_{21}', t_2)\,.\label{delpcom}
\eea
Of course, the extra $\td\delta$ terms on the right-hand side was not seen
in Schwarz's calculations, because he calculated the commutator only on
$M$, for which we know $\td\delta M=0$, but not on $X$.

   The conclusion from (\ref{delpcom}) is that if all the $\delta'$ 
transformations (\ref{schwtran}) are included in the symmetry algebra,
then it is necessary to extend the algebra further by including the
$\td\delta$ transformations too, in order to achieve closure.  
As may be seen from (\ref{schwtran}), Schwarz's
$\delta'$ transformations are themselves a combination of our $\delta$ and
$\td\delta$ transformations; in fact, one has
\be
\delta'(\ep_1) = \delta(\ep_1) -\td\delta(M_0^{-1} \ep_1^\dagger M_0)\,.
\label{deltapdef}
\ee
The upshot is that once one has extended Schwarz's transformations to
comprise not only $\delta'$ but also $\td\delta$, one has, equivalently,
extended to the full set of $\delta$ and $\td\delta$ transformations
that we considered in section \ref{kacmoodysec}.  These, as we showed,
generate the complete affine Kac-Moody extension $\hat\cG$ of the
original $\cG$ algebra.

   One can, alternatively, take a more restrictive viewpoint, which is
effectively the one that was adopted by Schwarz in \cite{schwarz1}.  Namely,
the commutation relations (\ref{delpcom}) imply that it is only if 
either $\delta_1'$ or $\delta_2'$ is a level-0 transformation that the
$\td\delta$ transformations are generated.  (This follows from the fact
that the second term on the right-hand side of (\ref{delpcom}) is 
independent of $t_2$, and the third term is independent of $t_1$.)
Thus, we have
\bea
[\delta_\m'(\epsilon_1), \delta_\n'(\epsilon_2)]
&=&\delta^S_{\sst{(m+n)}}(\epsilon_{12})\,,\qquad\hbox{for}\qquad
m>0\,,\quad n>0\,,\\\
[ \delta_{\sst{(0)}}'(\epsilon_1), \delta_\n'(\epsilon_2)]
&=&\delta^S_\n(\epsilon_{12}) + \td\delta_\n(M_0^{-1}\epsilon_{12}^\dagger
M_0 + \epsilon_{21}')\,,\qquad n>0\,.
\eea
(We have taken $\delta_1'$ to be a level-0 transformation in the second
equation, for definiteness.)  One can therefore avoid generating any 
$\td\delta$ transformations if one restricts the level-0 transformations
in $\delta'$ to be such that
\be
M_0^{-1}\epsilon_1^\dagger M_0 + \epsilon_1=0\,.\label{herm2}
\ee
This equation is essentially the condition that $\epsilon$ should belong
to the denominator algebra $\cH$ of the coset model.  This is most 
immediately clear if one chooses, as one may, the ``gauge'' in which
$M_0=1$.  Equation (\ref{herm2}) then implies that $\ep$ is anti-Hermitean,
which is precisely the standard condition for it to lie in the denominator
algebra $\cH$.  If some
other gauge choice is made for $M_0$, then $\ep$ is again required to
be in the denominator algebra, in a basis conjugated by 
 $M_0$.  The upshot of this
discussion is that the necessity of including all the $\td\delta$ symmetries
as well in order to achieve closure of the algebra (\ref{delpcom}) can
be avoided if one truncates to that subset of the $\delta'$ transformations
in which the $\cK$ transformations at 0-level are omitted.

   This, therefore, accounts for the symmetry algebra that was found by
Schwarz in \cite{schwarz1}.  The full Kac-Moody
symmetry algebra $\hat\cG$ is generated by our $\delta$ and 
$\td\delta$ transformations, whilst Schwarz's subalgebra, 
which he denoted by $\hat\cG_H$, corresponds
to the transformations $\delta'$ given in (\ref{deltapdef}), with the
further restriction that at level-0 the $\cK$ transformations are omitted.
Omitting these particular transformations is precisely what is needed in
order to maintain a fixed boundary condition for $M_0$ (such as $M_0=1$).
In the gauge choice $M_0=1$, we see from (\ref{deltapdef}) that 
$\delta'(\ep)= \delta(\ep) \pm \td\delta(\ep)$, with the plus sign occurring
when $\ep$ lies in $\cH$ and the minus sign when $\ep$ lies in $\cK$.  The
generators ${J'_n}^i$ of the Schwarz subalgebra are therefore given in
terms of the Kac-Moody generators $J^i_n$ by
\bea
{J'_n}^i &=& J^i_n + J^i_{-n}\,,\qquad\hbox{for}\qquad
i\in {\cal H}\,,\nn\\
{J'_n}^i &=& J^i_n - J^i_{-n}\,,\qquad\hbox{for}\qquad
i\in {\cal K}\,.\label{schwtrunc}
\eea
One sees immediately that the level-0 generators ${J'_0}^i$ vanish
if $t$ lies in $\cK$.
It can easily be verified directly that the generators ${J'_n}^i$ form 
a closed subalgebra of the full Kac-Moody algebra (\ref{kacmoody2}).

   The Schwarz subalgebra of the Kac-Moody algebra can be interpreted as
follows.  By writing $X(t)$ as in (\ref{pathi}), a choice has been made
to set $X(t)=1$ at the point $x_0$ in the two-dimensional spacetime.
This can be viewed as a gauge-fixing that is achieved by using the 
$\td\delta$ transformations.  Furthermore, as we remarked below (\ref{yyy}),
$M_0$ must be held fixed, which is a further gauge fixing (of the 
original sigma-model fields), achieved by using the $\cK$ part of the
original $\cG$ Lie algebra transformations.  In other words, only the
$\cH$ part of the original $\cG$ symmetry survives.
If we wish instead to retain the full algebra
$\cG$ of original symmetries, then Schwarz's subalgebra will necessarily
have to be extended to the full Kac-Moody algebra $\hat\cG$.

   It is instructive to look at this truncated subalgebra in the
concrete example of the $SL(2,\R)/O(2)$ sigma model that we studied
in section \ref{sl2rsec}.  Especially, it is interesting to look
at the transformations of the original $SL(2,\R)$ symmetry and the
dual $\overline{SL(2,\R)}$ symmetry, to see which are retained and
which are truncated out in the subalgebra.  

   The combinations of Kac-Moody generators $J^i_n$ that lie in 
$\cK$ and in $\cH$ are given, respectively, by
\bea
\cK:&& J^0_n\,,\qquad (J^+_n + J^-_n)\,,\nn\\
&&\nn\\
\cH:&& (J^+_n - J^-_n)\,.
\eea
It then follows from (\ref{schwtrunc}) that the generators ${J'_n}^i$
that are retained in the truncated algebra of \cite{schwarz1} are
\bea
\cK:&& {J'_n}^{\sst{(1)}}=J^0_n -J^0_{-n}\,,\qquad 
        {J'_n}^{\sst{(2)}}=J^+_n + J^-_n - J^+_{-n} - J^-_{-n}\,,\nn\\
&&\nn\\
\cH:&& {J'_n}^{\sst{(3)}}=J^+_n - J^-_n + J^+_{-n} - J^-_{-n}\,.
\label{schwtrunc2}
\eea

   Since the $SL(2,\R)$
transformations correspond to the Kac-Moody generators $J^i_0$, and the
$\overline{SL(2,\R)}$ transformations correspond to the generators
$J^+_{-1}$, $J^0_0$ and $J^-_1$, it suffices to consider just the
levels $m=0$ and $m=1$ in (\ref{schwtrunc2}).  These give the four following
non-vanishing generators:
\bea
n=0:&& {J'_0}^{\sst{(3)}}=2(J^+_0 - J^-_0)\,,\nn\\
&&\nn\\
n=1:&& {J'_1}^{\sst{(1)}}=J^0_1 -J^0_{-1}\,,\nn\\
    && {J'_1}^{\sst{(2)}}=J^+_1 + J^-_1 - J^+_{-1} - J^-_{-1}\,,\nn\\
    && {J'_1}^{\sst{(3)}}=J^+_1 - J^-_1 + J^+_{-1} - J^-_{-1}\,.
\eea
We see that just two of the five inequivalent transformations 
in $SL(2,\R)$ and $\overline{SL(2,\R)}$ are retained within the
truncated algebra:
\be
{J'_0}^{\sst{(3)}} \leftrightarrow (\ep^+-\ep^-)\,,\qquad
  ({J'_1}^{\sst{(2)}}-{J'_1}^{\sst{(3)}}) \leftrightarrow
     (\bep^+ -\bep^-)\,.
\ee
Thus, the infinite-dimensional subalgebra of the full Kac-Moody algebra
that is retained in the truncation (\ref{schwtrunc}) omits not only the
$\cK$ generators in the original $SL(2,\R)$, but also the $\cK$ generators
in the dual symmetry algebra $\overline{SL(2,\R)}$.  If one wants to 
have a symmetry algebra that at least contains all the 
generators of the original and the dual $SL(2,\R)$ algebras then, as we
showed in section \ref{gerochsec}, this will necessarily be the full Kac-Moody 
algebra.
    
\section{The Virasoro-type Symmetry and the Schwarz Approach}\label{Vapp}

   In section \ref{Virasorosubsec} we obtained a Virasoro-like symmetry
of the symmetric-space sigma models, with generators $K_n$ satisfying the
algebra (\ref{Valg2}).  Our construction was closely related to one given
in \cite{schwarz1} but there were significant differences, which we shall
elaborate on here.  

   The first respect in which our discussion diverges from that in 
\cite{schwarz1} is that in that paper, the quantity $\xi(t)$ appearing in
the our transformation $\delta^V(t) \cV= \cV \xi(t)$ (see (\ref{vsym}))
is replaced by
\be
\td\xi(t)= (t^2-1) \dot X(t) X(t)^{-1} + {\cal I}\,,\label{schwarzxi}
\ee
where 
\be
{\cal I} = \int {*A}\,.
\ee
One can see from the path-ordered integral expression (\ref{pathi}) for
$X(t)$ that 
\be
X(t)= 1 + t\, \int{*A} + {\cal O}(t^2)\,,
\ee
and so in fact ${\cal I}= \dot X(0) = \dot X(0) X(0)^{-1}$. Thus from
(\ref{schwarzxi}) we see that Schwarz's $\td\xi$ and our $\xi$ are related by
\be
\td\xi(t) = \fft{1-t^2}{t}\, \xi(t)-\left[\fft{1-t^2}{t}\, \xi(t)\right]_{t=0}
\,.\label{tdxi}
\ee
Thus the lowest mode in our transformation is excluded in the PCM analysis
in \cite{schwarz1}.

   The lowest mode had to be excluded in \cite{schwarz1} for
the principal chiral model, as opposed to the symmetric-space
sigma model, in order to ensure that the transformation was a 
symmetry of the equations of motion.  In brief, the transformation of $A$ under 
$\xi$ (defined as in (\ref{vsym})) in the PCM case is simply $\delta^V A=
D\xi$, rather than (\ref{delVA}) of the SSM case, and so using (\ref{DXX}) 
one finds
\be
\delta^V A= {*d}(t^{-1} \xi) + \fft1{1-t^2}\,  A + \fft{t}{1-t^2}\, {*A}\,.
\ee
This means that $d{*\delta^V A}= t/(1-t^2)dA$, 
and so the equation of motion $d{*A}=0$
is not preserved.  However, if the lowest-order term in
$\delta^V(t)$ is subtracted out, as is done in (\ref{tdxi}), then the
resulting transformation $\td\delta^V$ {\it does} give a symmetry.

   Although Schwarz carried over the assumption that the lowest mode 
should also be subtracted out when he then considered the SSM case, it is
actually no longer necessary to do so, as we explained in section
\ref{Virasorosubsec}.  As we showed there, with the transformation 
$\delta^V A$ now given by (\ref{delVA}), one finds using (\ref{DXX}) and
(\ref{DMXX}) that the contributions to $\delta^V A$ of the form ${*A}$
coming from the two terms in (\ref{delVA}) cancel out, and so 
$d{*\delta^V A}=0$ automatically, without the need to subtract the lowest mode
term.    The upshot is that the set of Virasoro-like symmetries that
we find for the symmetric-space sigma models is actually larger then the
set obtained by Schwarz in \cite{schwarz1}, by virtue of the inclusion of
the lowest mode in $\delta^V(t)$.

   A second difference between our results and those in \cite{schwarz1} 
is concerned with the precise form of the Virasoro-like algebra in the
two cases.  We were able to make a convenient choice of $-t$ as the
prefactor of $\dot X X^{-1}$ in (\ref{vsym}) which gave the 
algebra in the form (\ref{Valg2}), which is very close in structure to the
Virasoro algebra.  On the other hand, in \cite{schwarz1} the choice
of $t$-dependent prefactor was apparently constrained by certain 
requirements of matching between left and right acting transformations
on the group manifold of the PCM (a consideration that does not apply
in the SSM case).  This led to the choice of $(t^2-1)$
prefactor that was made in \cite{schwarz1}, and this in turn led
to the rather different algebra
\be
[K_m,K_n]= (m-n)(K_{m+n+1}- K_{m+n-1})
\ee
for the PCM case.

\end{document}